\documentclass{aa}  

\usepackage{graphicx}
\usepackage{txfonts}
\usepackage{multirow}
\usepackage{booktabs}
\usepackage{amsmath}
\usepackage{amssymb}
\usepackage{hyperref}
\usepackage{orcidlink}

\begin{document}

   \title{Pulsar Timing methods for evaluating dispersion measure time-series}

   \subtitle{}

   \author{F. Iraci\orcidlink{0009-0001-0068-4727}\inst{1}\fnmsep\inst{2}\thanks{\email{francesco.iraci@inaf.it}}
          \and
          A.~Chalumeau\orcidlink{https://orcid.org/0000-0003-2111-1001}\inst{3,4}
          \and
          C.~Tiburzi\orcidlink{0000-0001-66-51-4811}\inst{1}
          \and
          J.~P.~W.~Verbiest\orcidlink{0000-0002-4088-896X}\inst{5}
          \and
          A.~Possenti\orcidlink{0000-0001-5902-3731}\inst{1}
          \and
          G.~M.~Shaifullah\orcidlink{0000-0002-8452-4834}\inst{3,4} \and
          S.~C.~Susarla\orcidlink{0000-0003-4332-8201}\inst{6}
          \and
          M.~A.~Krishnakumar\orcidlink{0000-0003-4528-2745}\inst{7,8,9}
          \and
          M.~T.~Lam \orcidlink{0000-0003-0721-651X}\inst{10,11,12}
          \and
          H.~T.~Cromartie \orcidlink{0000-0002-6039-692X}\inst{13}
          \and
          M.~Kerr \orcidlink{0000-0002-0893-4073}\inst{14}
          \and
          Jean-Mathias~Grie{\ss}meier\orcidlink{0000-0003-3362-7996}\inst{15,16}
          }

   \institute{
  INAF - Osservatorio Astronomico di Cagliari, via della Scienza 5, 09047 Selargius (CA), Italy
  \and
  Dipartimento di Fisica, Università di Cagliari, Cittadella Universitaria, I-09042 Monserrato (CA), Italy
 \and
  Dipartimento di Fisica ``G. Occhialini'', Università degli Studi di Milano-Bicocca, Piazza della Scienza 3, I-20126 Milano, Italy
  \and
  INFN, Sezione di Milano-Bicocca, Piazza della Scienza 3, I-20126
Milano, Italy
  \and
  Florida Space Institute, University of Central Florida, 12354 Research Parkway, Orlando, FL 32826, USA
  \and
  Physics, School of Natural Sciences, Ollscoil na Gaillimhe --- University of Galway, University Road, Galway, H91 TK33, Ireland
  \and
  Max-Planck-Institut f{\"u}r Radioastronomie, Auf dem H{\"u}gel 69, 53121 Bonn, Germany\label{mpifr}
  \and
  Fakult{\"a}t f{\"u}r Physik, Universit{\"a}t Bielefeld, Postfach 100131, 33501 Bielefeld, Germany\label{unibi}
  \and
  National Centre for Radio Astrophysics, Pune University Campus, Pune 411007, India \label{ncra}
  \and
  SETI Institute, 339 N Bernardo Ave Suite 200, Mountain View, CA 94043, USA
  \and
  School of Physics and Astronomy, Rochester Institute of Technology, Rochester, NY 14623, USA
  \and
  Laboratory for Multiwavelength Astrophysics, Rochester Institute of Technology, Rochester, NY 14623, USA
  \and
  National Research Council Postdoctoral Associate, National Academy of Sciences, Washington, DC 20001, USA resident at Naval Research Laboratory, Washington, DC 20375, USA
  \and
  Space Science Division, Naval Research Laboratory, Washington, DC 20375--5352, USA
  \and
  LPC2E - Universit\'{e} d'Orl\'{e}ans /  CNRS, 45071 Orl\'{e}ans cedex 2, France
  \and
  Observatoire Radioastronomique de Nan\c{c}ay (ORN), Observatoire de Paris, Universit\'{e} PSL, Univ Orl\'{e}ans, CNRS, 18330 Nan\c{c}ay, France
}

   \date{}

 
  \abstract
   {Radio pulsars can be used for many studies, including the investigation of the ionised interstellar medium and the solar wind via their dispersive effects. These phenomena affect the high-precision timing of pulsars and are among the main sources of noise in experiments searching for low-frequency gravitational waves in pulsar data.} 
   {In this paper, we compare the functionality and reliability of three commonly used schemes to measure temporal variations in interstellar propagation effects in pulsar-timing data.}
   {We carry out extensive simulations at low observing frequencies (100-200~MHz) by injecting long-term correlated noise processes with power-law spectra and white noise, to evaluate the robustness, accuracy and precision of the following three mitigation methods: epoch-wise (\textbf{EW}) measurements of interstellar dispersion; the \textbf{DMX} method of simultaneous, piece-wise fits to interstellar dispersion; and \textbf{DM GP}, which models dispersion variations through Gaussian processes using a Bayesian analysis method. We then evaluate how reliably the input signals are reconstructed and how the various methods react to the presence of achromatic long-period noise.}
   {All the methods perform well, provided the achromatic long-period noise is modeled for \textbf{DMX} and \textbf{DM GP}.
    The most precise method is \textbf{DM GP}, followed by \textbf{DMX} and \textbf{EW}, while the most accurate is \textbf{EW}, followed by \textbf{DMX} and \textbf{DM GP}.
    We also test different scenarios including simulations of L-band ToAs and realistic DM injection, with no significant variation in the obtained results.
    }
   {Given the nature of our simulations and our scope, we deem that \textbf{EW} is the most reliable method to study the Galactic ionized media. Follow-up works should be conducted to confirm this result via more realistic simulations.
   We note that \textbf{DM GP} and \textbf{DMX} seem to be the most performing techniques in removing long-term correlated noise, and hence for gravitational wave studies. However, full simulations of pulsar timing array experiments are needed to support this interpretation.}

   \maketitle

\section{Introduction}

Pulsars are highly magnetized, rapidly rotating neutron stars that emit beams of radiation, mainly in the radio band, from their magnetic poles. 
Therefore, pulsars are visible as periodic sources to terrestrial observers.
The extremely regular rotation of a pulsar allows to predict the time of arrival (ToA) of its radiation at a radio observatory and to extract from these data information about the pulsar itself, its environment and perturbing effects through the pulsar timing procedure \citep[e.g.,][]{Kramer2005handbook}.
The model which contains the parameters necessary to describe the signal propagation, usually referred to as timing model, is used to calculate expected ToAs that are then compared with the observed ones to create and analyze the timing residuals (the difference between the observed ToAs and the ones predicted by the timing model). 
The parameters in the timing model are iteratively fit to minimize the timing residuals and to achieve a more and more precise description of the analyzed pulsars. 
The shorter the spin period and the brighter the pulsar, the smaller its weighted root-mean-square (RMS) timing residuals and the more precisely determined are the timing-model parameters.
The most rapidly rotating pulsars are known as "millisecond pulsars" \citep{backer1982millisecond} because they have spin periods on the order of milliseconds. 
The technique of pulsar timing opens up the study not only of the physical characteristics of pulsars, but also other effects, such as the propagation effects through ionized media \citep{Rickett1990, Foster1990}; the most conspicuous of which is dispersion.\\
Dispersion is a phenomenon caused by the dependency of a medium's refractive index $\mu$ on the frequency $\nu$ of the propagating radiation and on its own free-electron density $n_e$. In particular:
\begin{equation}
    \mu = \sqrt{\left ( 1 - \frac{f_{\rm p}}{\nu} \right )}
\end{equation}
with $f_p$, the plasma frequency, defined as:
\begin{equation}
    f_{\rm p} = \sqrt{\frac{e^2 n_{\rm e}}{\pi m_{\rm e}}}
\end{equation}
where $e$ and $m_e$ are respectively the electron electric charge and mass. The net effect of these two dependencies is an absolute delay in the arrival time of the radiation bundle after propagating through an ionized medium (such as the interstellar medium, which hosts an ionized component, or the solar wind), and a relative delay among the radiation at different frequencies within that bundle. 
In particular, we have that the delay $\Delta t$ for radiation with a frequency $\nu$ is approximately (assuming $f_{\rm p} \ll \nu$):
\begin{equation}\label{eq:deltat}
    \Delta t = \mathcal{D} \frac{DM}{\nu^2}
\end{equation}
where $\mathcal{D}$ is the dispersion constant \citep{Kramer2005handbook},
and the $DM$ parameter is the dispersion measure, defined as:
\begin{equation}
    DM = \int_{LoS} n_{\rm e} {\rm d}l
\end{equation}
where the integral is performed along the line of sight (LoS) and the units are $\mathrm{pc}\,\mathrm{cm^{-3}}$.
The DM of a pulsar is one of the parameters included in its timing model.

Equation~\eqref{eq:deltat} can be rewritten following \citet{vs18}:
\begin{equation}\label{eq: js18}
    \Delta t = 4.15 \times 10^3 DM \left(\frac{2B}{\nu_c^2}\right)
\end{equation} 
where $B$ is the fractional bandwidth $(\nu_2-\nu_1)/{\nu_c}$ and $\nu_c$ is the central frequency of the observing band. 

From these equations it is easy to infer that i) the DM is a direct measure of the free-electron content along a given line of sight and ii) large delays between radiation emitted at different frequencies of the same bundle are shown over low central observing frequencies and/or large fractional bandwidths. 
Note that, in practice, an absolute value of DM can never be achieved due to the evolution of pulsar profiles with the observing frequency \citep[e.g.,][]{hsh+12}. 
To complicate matters further, pulsars are high-velocity objects, following the kick that they receive during the supernova event \citep{hllk05}. 
This implies that they move rapidly across the sky, and hence the LoS varies significantly as well. 
In doing so, different parts of the ionized interstellar medium (IISM) are crossed, with different associated values for $n_{\rm e}$, and hence the value of DM changes with time \citep{Lam2016}.
simple polynomial models of DM time evolution
This is why simple polynomial models of DM time evolution are often included in the timing model.
Nevertheless, the turbulence in the IISM cannot be correctly described via derivatives only, and hence there are a large number of studies in literature reporting on the time-series of DM variations and their connection with the physics of plasma in the Galaxy. 

Recently, \citet{Donner2020} reported the DM variations for 36 millisecond pulsars observed for about 7 years at low-radio frequency (100-200~MHz) with the LOFAR (the LOw Frequency ARray, see \citealt{vanHarleem2013}) interferometer in which they reached a very low median DM uncertainty, of the order of $10^{-5}$~pc/cm$^{3}$, for a significant fraction of the sample.
Besides, DM variations were detected when the median DM uncertainty was lower than a few in $10^{-4}~$pc/cm$^{3}$.
\citet{KK2021} presented the 1-year long DM time series of four millisecond pulsars observed with uGMRT using BAND3 (400-500~MHz) and BAND5 (1360-1460~MHz).
They showed that the DM precision improves up to $10^{-4}$~pc/cm$^{3}$ when combining data from both of the bands.
\citet{NG15timing} and \citet{Jones2017} reported on the DM variations of up to 68 millisecond pulsars with mainly the Green Bank (722-1885~MHz) and the Arecibo (302-2400~MHz) radio telescopes, highlighting on their monotonic trends and the presence of annual IISM signatures for almost 20 of them.
Lastly, \citet{Keith2024} calculated the DM time series for almost 600 long-period pulsars observed with the MeerKAT telescope at L-band (896-1671~MHz) and a large fractional bandwidth, reporting a broad linear correlation between the DM and its first derivative.

While DM variations allow for the investigation of Galactic plasma, when they are unaccounted for may become a nuisance in other experiments such as Pulsar Timing Arrays \citep[PTAs,][]{Tiburzi2018reviewPTA, Verbiest_2021, Taylor2021}, that search for low-frequency gravitational waves \citep[GWs,][]{Maggiore2018} in pulsar data.
In particular, GWs are expected to cause space- and time-correlated, long-term perturbations in pulsar timing residuals, usually characterized by a steep, red power spectrum \citep{PHI2001}. 
PTAs search for these effects by cross-correlating the timing residuals of pairs of selected pulsars, to identify the predicted signature following the so-called Hellings \& Downs curve in the case of an isotropic GW background \citep{HDcurve1983}. 
One of the most challenging parts of PTA analyses is the characterization of signals that are not GWs but that also induce long-term, time-correlated structures in the timing residuals \citep[see, e.g.,][]{vs18}, i.e., that are sources of red noise.
The red-noise processes with the highest amplitude have two main contributions. 
The first is often called timing noise, or spin noise, and refers to rotational instabilities of the targeted pulsar (\citealt{Hobbs2004}).
The second is the aforementioned time variability in the amount of interstellar dispersion affecting the pulsar radiation (hereafter referred to as DM noise).
PTAs have historically been focused on high frequency bands ($\sim1-3~\mathrm{GHz}$), small fractional bandwidths and/or asynchronous timing observations at different frequency bands \citep[see][where they demonstrated that multi-frequency asynchronous observations will never reduce the timing error due to DM noise below $10\mathrm{ns}$]{Lam2015}.
Therefore, while the DM noise has indeed an impact on the GW search, PTA data have a poor sensitivity towards it \citep[see also Figure 6 in][]{vs18}.
This is why PTAs are starting to use observing campaigns obtained with low-frequency, large-bandwidth facilities such as LOFAR and NenuFAR (the New Extension in Nançay Upgrading LOFAR, see \citealt{NenuFAR2012}), see e.g. \citet{Donner2020, Tiburzi2021, Bondonneau2021}.




In this paper we aim to assess the performances in terms of precision and accuracy of three methods to calculate DM variations, that have been used to study Galactic plasma or to model the DM noise impact in PTA data, based on a comprehensive series of simulations.
In Section~\ref{sec:2} we describe these three approaches and the simulations.
In Section~\ref{sec:3} we report the simulation results, which are discussed in Section~\ref{sec:4}. 
Finally, in Section~\ref{sec:5} we draw our conclusions.

We do not assess the impact of GWs on the DM recoveries methods and also we do not check the impact of these schemes on the timing model parameters.
These aspects will be tested in future works.

We stress that there are also other effects that might affect pulsar timing at low radio frequencies.
With LOFAR observations the ToAs are generated using a frequency-resolved template \citep[see, e.g.,][]{Donner2020, Tiburzi2021} and hence interstellar scintillation is not affecting the timing solution.
Pulse broadening and scattering variations are two other features that might be present in low frequency observations.
In this paper we do not take into account these effects and we simulate a more commonly occurring scenario, without frequency-dependent contributions other than DM variations.

\section{Methods / Simulations and analysis}\label{sec:2}
\subsection{Simulations}

    To compare the effectiveness of the various PTA  methods to calculate and account for DM variations we evaluate their application on simulated data, which are produced using the \texttt{libstempo} software package \citep{Libstempo2020}.

\subsubsection{Simulations with \texttt{libstempo}} \label{subsubsec:libstempo}

    \texttt{libstempo} is a python wrapper of the \texttt{TEMPO2} software package \citep{Tempo2} which allows the use of all of the \texttt{TEMPO2} functionalities within a python-based environment. 
    In particular, we exploit the  \texttt{libstempo.toasim.fake\_pulsar} library to simulate 10 narrow-band ToAs per observing epoch in the most commonly used LOFAR frequency band for pulsar observations, between $100$ and $200 \, \mathrm{MHz}$ with $5~\mu\mathrm{s}$ fixed ToA template-fitting error bars. The simulated observing epochs have a regular fortnightly cadence, and cover a time span of $T_{span}=3000$ days.
    To build simulations as close as possible to the typical PTA dataset, we inject stochastic noise in the form of white noise, achromatic (i.e., radio frequency independent) red noise and DM variations.

    White noise (WN) reflects instrumental errors and instrumental sensitivity, as well as intrinsic pulse jitter.
    We model it by considering two parameters: 
    EFAC which is a multiplicative factor that takes into account ToA measurement errors;
    EQUAD which is added in quadrature and accounts for any other white noise that may be given by profile variations and possible systematic errors.
    The final ToA uncertainty is then:
    \begin{equation}
    \label{eq: WN}
        \sigma_{\text{ToA}} = \sqrt{(\text{EFAC} \cdot \sigma_{\text{temp}})^2 + \text{EQUAD}^2},
    \end{equation}
    where $\sigma_{\text{temp}}$ is the template-fitting error and it is given as an input to our simulations\footnote{We use the same convention of \citet{EPTA2023paperII} for $\sigma_{ToA}$, however some PTA collaborations use the \texttt{TEMPO2} definition \citep[see][]{vlh+16}.}.
    
    
    Achromatic red noise (RN) and DM variations are both time correlated noise processes which are modeled with a Fourier basis of $N_f$ coefficients and a power-law power spectrum \citep{EPTA2023paperII}.
    We distinguish between chromatic and achromatic noise processes using a chromatic index $\alpha$ respectively equal to 2 and 0.
    The time delay induced on a ToA with frequency $\nu$ at an epoch $t$ is then:
    \begin{equation}
        t_{\text{delay}}(t) = \sum_{i=1}^{N_f}\sqrt{\mathcal{P}_i} F_i \left( \frac{\nu}{\nu_{\text{ref}}} \right)^{-\alpha}\, .
        \label{eq: Fourier sum injection}
    \end{equation}
    $\mathcal{P}_i$ is the powerlaw power spectral density with hyperparameters $\Hat{A}$, the normalized amplitude at the frequency of $1~\mathrm{yr^{-1}}$, and $\gamma$, the spectral index. 
    It is written as follow: 
    \begin{equation}
        \mathcal{P}_i = \frac{\Hat{A}_{n}^2}{12\pi^2} \frac{\text{yr}^3}{T_{span}} \left( \frac{f_i}{1\text{yr}^{-1}} \right)^{-\gamma_{n}}
        \label{Power-law injection}
    \end{equation}
    with $n=\text{RN or DM}$ and we will refer to the amplitude as $A_{n} = \log_{10}(\Hat{A}_{n})$.
    $F$ is the matrix of the cosine functions at each Fourier frequency $f_i = i/N_f$ and time $t$: 
    \begin{equation}
        F_i(t) = \sqrt{2}\,\cos\left( 2\pi f_i t + \phi \right),
        \label{cosine functions injection}
    \end{equation}
    with $\phi$ being a random phase drawn from $\mathcal{U}(0,2\pi)$.
    The radio frequency is $\nu$ and the reference frequency $\nu_{ref}$ is set to 1.4~GHz.
    The time series of the injected noise signal will then be a sum over a finite number $N_f$ of cosine functions (see second and third row of Figure~\ref{fig: sim injection plot} for an example).
    For both RN and DM noise we consider $N_f = 30$ components. 
    This corresponds to a frequency of about $1.15\times\,10^{-7}~\mathrm{Hz}$ which is close to the Nyquist one that, given our 14 days of cadence, is at around $2\times10^{-7}~\mathrm{Hz}$.

\subsubsection{Simulation parameters}

    Table~\ref{tab: Parameters} shows all the parameters adopted for our simulations, with $\sigma_{\text{temp}}$, EFAC and EQUAD being fixed respectively to $5~\mu\mathrm{s}$, 1.2 and 2~$\mu\mathrm{s}$ for all the cases.
    \begin{table}[]
    \caption{List of the input noise parameter values adopted for our simulations.
    }
    \centering
    \begin{tabular}{ccc}
    \toprule
    \multicolumn{3}{c}{EFAC = 1.2       \, EQUAD = $2\cdot10^{-6}~\mathrm{s}$}\\ 
    \small $(\gamma_{RN}=3.7$; $\gamma_{DM}=8/3$) & \text{or} & \small $(\gamma_{RN}=2.5$; $\gamma_{DM}=3.2)$ \\
    \bottomrule
    \multicolumn{1}{c}{\begin{tabular}[c]{@{}c@{}}$A_{RN}$ = -12.3\\ $A_{DM}$= -12.6\end{tabular}} & \multicolumn{1}{c}{\begin{tabular}[c]{@{}c@{}}$A_{RN}$ = -12.3\\ $A_{DM}$= -13.6\end{tabular}} & \multicolumn{1}{c}{\begin{tabular}[c]{@{}c@{}}$A_{RN}$ = -12.3\\ $A_{DM}$= -14.6\end{tabular}} \\ \hline
    \multicolumn{1}{c}{\begin{tabular}[c]{@{}c@{}}$A_{RN}$ = -13.3\\ $A_{DM}$= -12.6\end{tabular}} & \multicolumn{1}{c}{\begin{tabular}[c]{@{}c@{}}$A_{RN}$ = -13.3\\ $A_{DM}$= -13.6\end{tabular}} & \multicolumn{1}{c}{\begin{tabular}[c]{@{}c@{}}$A_{RN}$ = -13.3\\ $A_{DM}$= -14.6\end{tabular}} \\ \hline
    \multicolumn{1}{c}{\begin{tabular}[c]{@{}c@{}}$A_{RN}$ = -14.3\\ $A_{DM}$= -12.6\end{tabular}} & \multicolumn{1}{c}{\begin{tabular}[c]{@{}c@{}}$A_{RN}$ = -14.3\\ $A_{DM}$= -13.6\end{tabular}} & \multicolumn{1}{c}{\begin{tabular}[c]{@{}c@{}}$A_{RN}$ = -14.3\\ $A_{DM}$= -14.6\end{tabular}} \\ 
    \bottomrule
    \end{tabular}

    \tablefoot{
    The WN parameters EFAC and EQUAD are fixed in all the realisations respectively to $1.2$ and $2~\mu\mathrm{s}$ and $\sigma_{\text{temp}}=5~\mu\mathrm{s}$.
    We use two different pairs of spectral indices: the first and more realistic with a steeper RN ($\gamma_{RN}=3.7$, $\gamma_{DM}=8/3$);
    the second and opposite case is produced with a flatter RN spectrum ($\gamma_{RN}=2.5$, $\gamma_{DM}=3.2$).
    For each pair we use 9 different combinations of ($A_{RN}$, $A_{DM}$) as reported in each cell.
    }
    \label{tab: Parameters}
    \end{table}
    Concerning the noise spectral indices, we first assume a realistic spectral index for DM variations from the Kolmogorov turbulence theory \citep{ARS95} ($\gamma_{DM}=8/3$) and a steeper one for the RN ($\gamma_{RN}=3.7$).
    Secondly, we consider a different pair of spectral indices: $\gamma_{DM}=3.2$ and $\gamma_{RN}=2.5$. 
    For each pair of spectral indices we have studied three fixed values of amplitude of the power spectrum for both of the noise processes, producing a total of nine datasets per pair of spectral indices.
    $A_{RN}$ and $A_{DM}$ have been chosen within the range of values reported by  \cite{EPTA2023paperII}.

    For each specific set of $A_{RN}$ and $A_{DM}$ and pair of spectral indices, we generate a total of 100 timing residuals realizations. In each realization, the seed for the injected RN is kept constant to maintain the same RN signature across all timing residuals, while the DM noise seed varies with each iteration, resulting in distinct DM time series for all 100 realizations.

\begin{figure*}
        \centering
        \includegraphics[width=\textwidth]{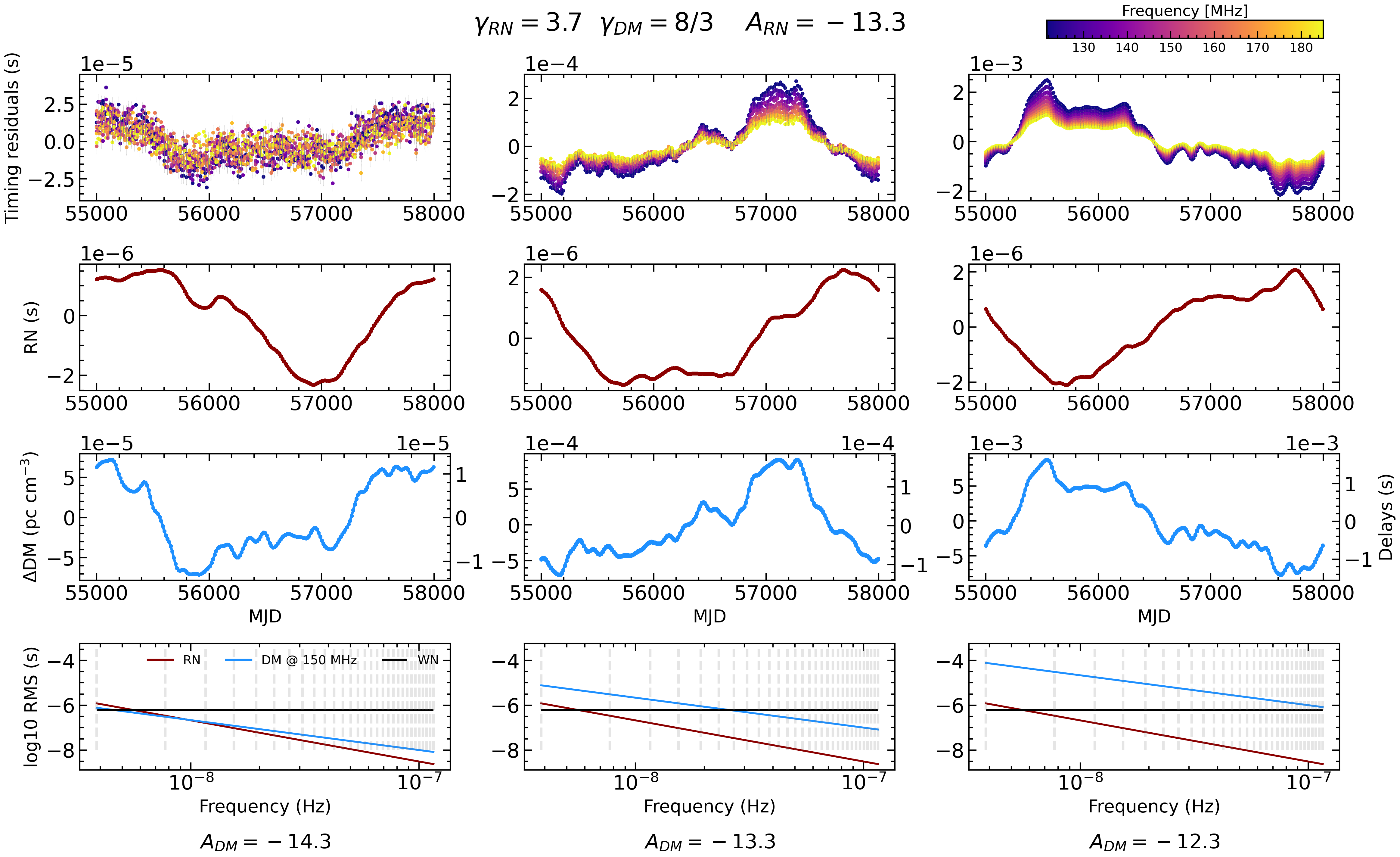}
        \caption{Simulations carried out with injected values of $\gamma_{RN}$, $A_{RN}$ and $\gamma_{DM}$ of, respectively, $3.7$, $-13.3$ and $8/3$.
        The columns are associated with increasing values of $A_{DM}$, left to right (as indicated at the bottom).
        First row: One out of the 100 simulated timing residuals with a colormap associated to the observing frequency (as reported in the top right corner).
        Second row: Injected RN signal.
        Third row: One out of the 100 different injections of the DM time series. On the right y-axis we report the corresponding time delay at the reference frequency of $150~\mathrm{MHz}$.
        Fourth row: Power spectra of the injected noise processes: WN (black), RN (red), DM variations at LOFAR central frequency of the band: $\sim150$MHz (light blue).
        Grey dashed vertical lines are the 30 Fourier frequency bins.
}
        \label{fig: sim injection plot}
    \end{figure*}

    Figure \ref{fig: sim injection plot} clarifies our simulation process by reporting three different scenarios, each of them with a different $A_{DM}$ that increases going from left to right.
    The first row reports the overall timing residuals; 
    the second row shows the delay due to the injected RN;
    the third row illustrates the DM variations introduced in our dataset.
    The last row represents the power spectrum of the noise processes involved in the three scenarios.
    Note that the first and third rows of each column reports only one of the 100 realizations that have been created for that set of parameters.

\subsection{DM recovery methods}
    
    Each generated series of timing residuals is analysed with the following methods.
    
\subsubsection{Epoch-wise DM}

    Epoch-wise DM \citep[\textbf{EW}, e.g. ][]{Donner19,TIB19} models a DM value and a constant offset $C$ via \texttt{TEMPO2} over the parsed ToAs for each simulated observing epoch.
    In particular the fit functional form is:
    \begin{equation}
        \Delta t = \frac{\text{DM}}{\nu^2} + C
        \label{eq: EW fit}
    \end{equation}
    where the parameters are the dispersion measure DM, and $C$, a constant offset that is supposed to absorb any un-accounted achromatic process.
    The observing frequency $\nu$ corresponds in our simulations to an array of 10 values in the LOFAR frequency range 110-190 $\mathrm{MHz}$ and $\Delta t$ is the time delay contained in the measured ToAs.
    This \textbf{EW} analysis returns a pair of DM and its corresponding 1$\sigma$ uncertainty given by the fit for each observation.
    The advantage of \textbf{EW} is that it simultaneously accounts for the DM and RN effects on the set of parsed ToAs, that corresponds to the ToA set for a single observing epoch. 
    Hence, we do not need to separately model the RN.
    
\subsubsection{DMX}

    \textbf{DMX} is a method developed by the NANOGrav collaboration \citep[e.g.,][~sec 4.1]{NG15timing} which performs a piece-wise constant fit of a DM value across an a-priori specified temporal window. 
    In our simulations, each window is such that it contains the 10 ToAs of a single observing epoch so that, at the end of the procedure, we obtain as many \textbf{DMX} parameters as the number of observing epochs.
    In this way \textbf{DMX} is extremely similar to Epoch-wise, with two important differences:
    it uses the \texttt{PINT} software \citep{PINT2021} suite to perform the fit, and it does not model the constant offset $C$.
    This means that fitting for the \textbf{DMX} parameters does not account for any RN process in the data and, in order to do so, the procedure to calculate an unbiased DM time series is divided in three steps:
    a first fit for the \textbf{DMX} parameters; a RN modeling round (while correcting for the initial \textbf{DMX} measures) with the \texttt{enterprise} software suite \citep{Enterprise2020}; then a second \textbf{DMX} fit.
    In the end, also \textbf{DMX} will provide a DM value, with the corresponding uncertainty, for each observation.
    
\subsubsection{DM GP}
\label{subsubsec: DM GP}
The EPTA and PPTA collaborations use a fully Bayesian-based noise-analysis approach carried out with the \texttt{enterprise} software suite, able to describe the WN with the EFAC and EQUAD parameters (described in Section~\ref{subsubsec:libstempo}), and the achromatic red noise (RN) and DM variations, which are modeled as stationary Gaussian Processes \citep[GPs, see details in appendix \ref{Appendix LaForge} and][]{vanHaasteren2014}, while the Bayesian inference is performed via the Markov-Chain Monte Carlo sampler \texttt{PTMCMCSampler} \citep{Ellis2017PTMCMC}. 
In this approach, the time delay induced on a ToA with radio frequency $\nu$ at an epoch $t$ is written as:
    \begin{equation}
        \delta t(t) = \sum_{i=1}^{N_f}\left[ a_i\,\sin(2\pi tf_i) + b_i\,\cos(2\pi t f_i)  \right]\left(\frac{\nu}{\nu_{\text{ref}}} \right)^{-\alpha}\, ,
        \label{eq: Fourier sum}
    \end{equation}
    where $\nu_{ref}$, $\alpha$ and $f_i$ are the same as in Equation \eqref{eq: Fourier sum injection}.
    The number of frequency components we use is the same of the injection: $N_f=30$ for both RN and DM.
    The weights $a_i$ and $b_i$ follow a multivariate Gaussian distribution with zero mean and covariance matrix $\Sigma_{jk}$ that in the frequency domain is given by:
    \begin{equation}
        \Sigma_{jk} = P_L(f_j)\frac{\delta_{jk}}{T_{span}},
    \end{equation}
    with $j,k=1...N_f$, the Kronecker delta $\delta_{jk}$ and the power spectrum $P_L$:
    \begin{equation}
        P_L(f) = \frac{A^2\, \text{yr}^3}{12\pi^2}\left(\frac{f}{1\text{yr}^{-1}}\right)^{-\gamma}.
    \end{equation}
By marginalizing over the $a_i$ and $b_i$ parameters, this analysis yields the posterior distributions for EFAC and EQUAD, and for the hyperparameters $A$ and $\gamma$ of RN and DM power spectra.
The posteriors describing the DM power-spectrum allow to reconstruct the DM noise via the \texttt{LaForge} software suite \citep{hazboun2020LaForge}, as a sum over a set of $N_f$ Fourier components as described in appendix~\ref{Appendix LaForge}. 
    %
    
    In particular,  we give as input to \texttt{LaForge} a large number $N$ of random posterior samples of the noise parameters, and we obtain as output $N$ different time-domain realizations of the time delays corresponding to the same noise process.
    These delays are then converted into DM values by inverting Equation \eqref{eq:deltat}, hence yielding a set of $N$ DM time series (i.e., a DM probability distribution for each epoch).
    In order to compare the \textbf{DM GP} output with the results of the other two methods we compute the mean DM and standard deviation $\sigma_{DM}$ at each epoch, hence obtaining one reference value per observation.   

\section{Results}\label{sec:3}
    
\begin{figure*}
        \centering
        \includegraphics[width=\textwidth]{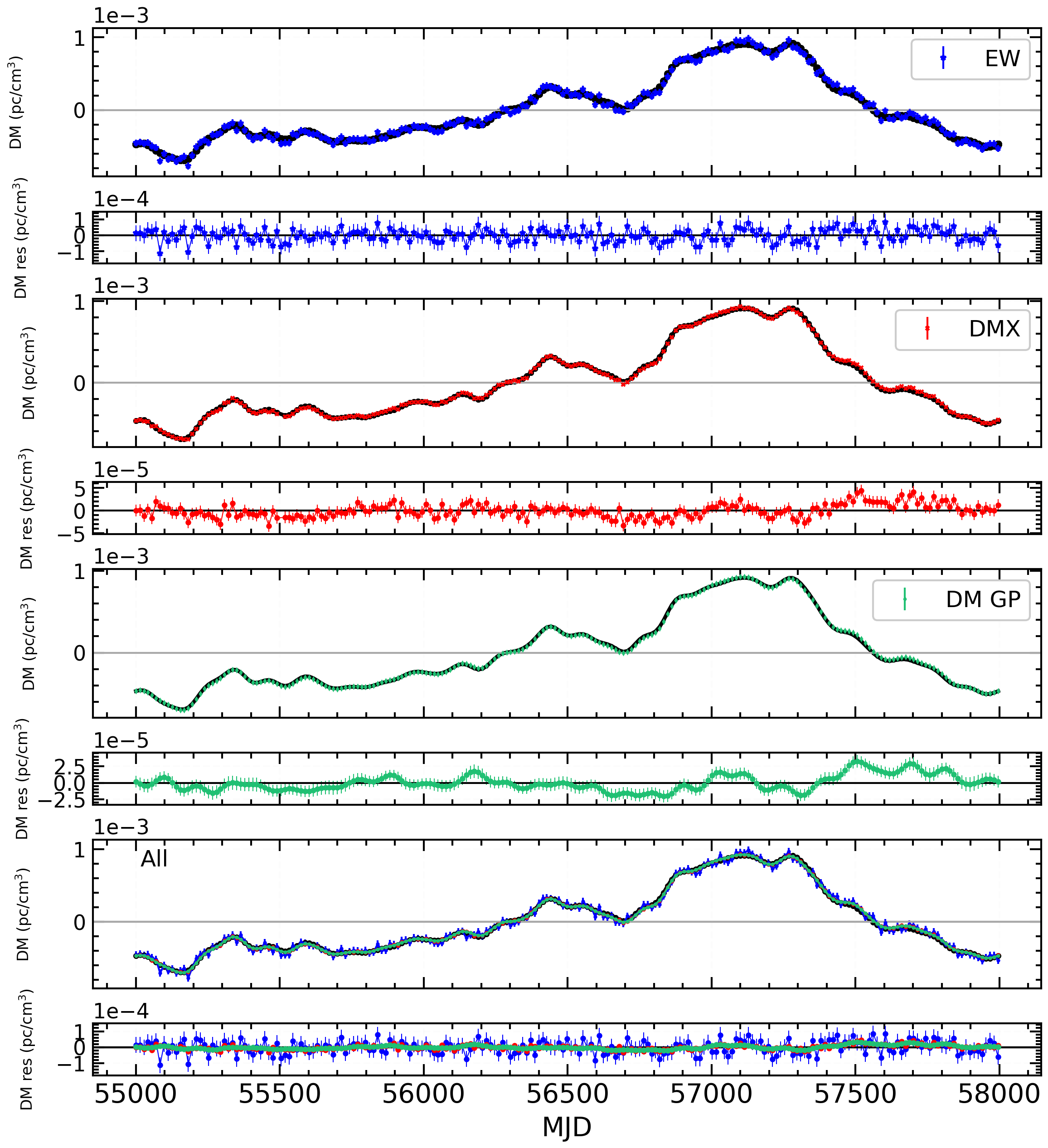}
        \caption{DM time series and DM residuals as recovered by \textbf{EW} (blue, 1st and 2nd panel), \textbf{DMX} (red, 3rd and 4th panel) and \textbf{DM GP} (green, 5th and 6th panel).
        The black line is the injected DM time series.
        In the last two panels we overplot, respectively, the DM time series and the DM residuals for the three methods.
        The DM and RN injected parameters for this realization are: $\gamma_{RN} = 3.7$, $\gamma_{DM} = 8/3$, $A_{RN} = -12.6$, $A_{DM} = -13.3$}
        \label{fig: DM recoveries}
    \end{figure*}

\begin{figure*}
        \centering
        \includegraphics[width=\textwidth]{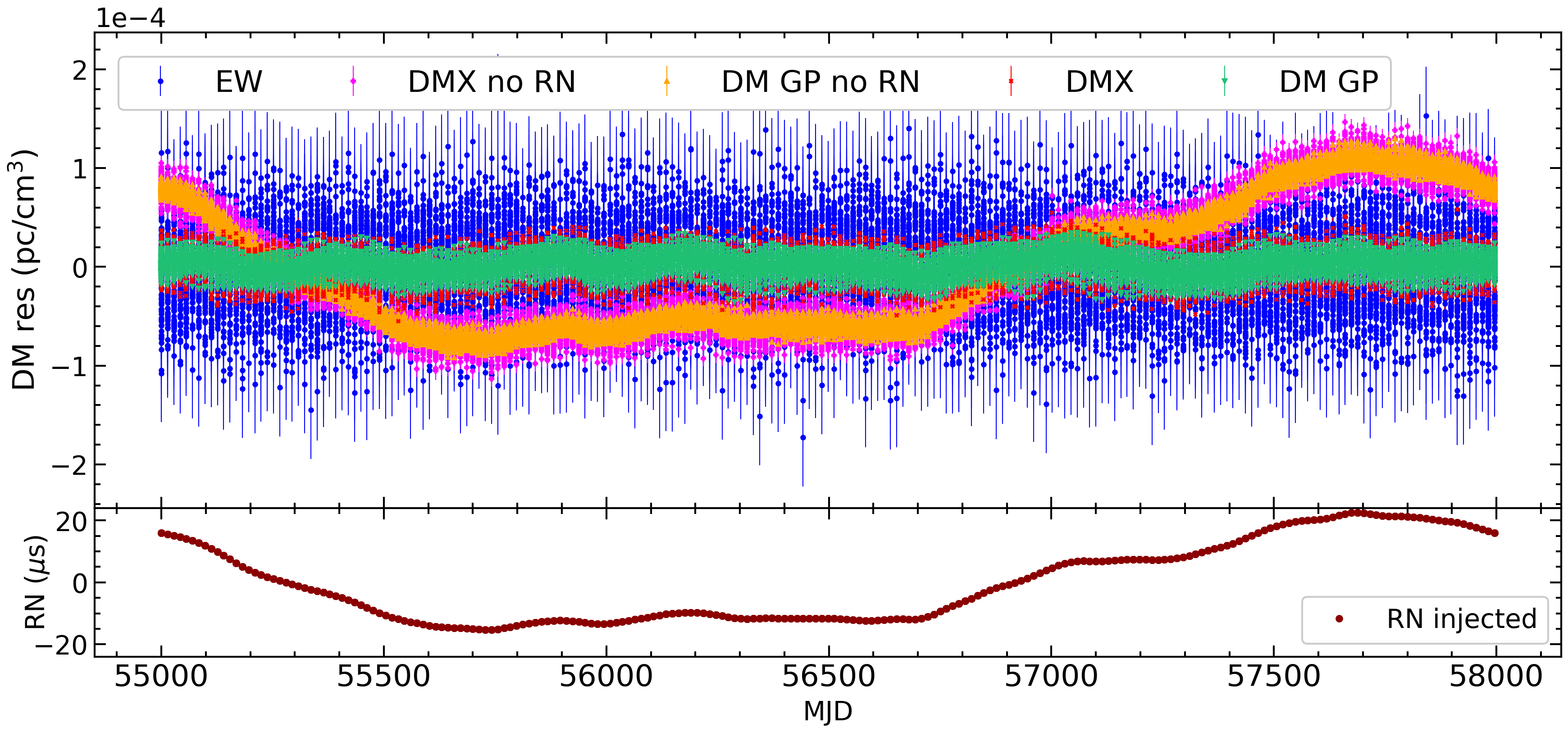}
        \caption{Top panel: DM residuals of the 100 realization for the five considered methods: \textbf{EW}(blue), \textbf{DMX no RN} (magenta), \textbf{DM GP no RN} (orange), \textbf{DMX} (red), \textbf{DM GP} (green) for set of input parameters: $\gamma_{RN} = 3.7$, $\gamma_{DM} = 8/3$, $A_{RN} = -12.6$, $A_{DM} = -13.3$.
        Bottom panel: RN signal injected (the same for all of the 100 realizations. See Section~\ref{sec:2})}
        \label{fig: DM residuals 100}
    \end{figure*}

We present here the results of the analysis with each method. A total of 3 main methods have been considered (\textbf{EW}, \textbf{DMX}, \textbf{DM GP}), and two additional ones in which the RN is not modeled (\textbf{DMX no RN}, \textbf{DM GP no RN})\footnote{RN modeling is an inherent part of \textbf{EW}, so no separate analysis is possible.}:
In Figure~\ref{fig: DM recoveries} are shown the DM time series recovered with the three main methods on the same noise realization accompanied by the corresponding DM residuals obtained by subtracting the recovered and the injected DMs. 
    An example of the DM residuals that can be obtained by combining together all the 100 realizations generated per each simulation is shown in Figure~\ref{fig: DM residuals 100}, with the bottom panel representing the RN signal injected in all realizations.
    In Tables~\ref{tab: errors RN -11/3, DM -8/3} and \ref{tab: errors RN -2.5, DM -3.2} we report the mean error on the DM estimates for each method and for each set of parameters.
    Figures~\ref{fig: DM residuals hist gamma 3.6}~and~\ref{fig: DM residuals hist gamma 2.5} report DM residual histograms summarizing the results obtained for all the simulations per pair of spectral indices. 
    In particular, the values reported in the histograms are obtained by normalizing the DM residuals by their errorbars, in order to have insights about their Gaussianity by fitting them via a Gaussian function. The $\chi^2$ of these fits are given in Tables~\ref{tab: chi2 RN -11/3, DM -8/3} and \ref{tab: chi2 RN -2.5, DM -3.2}.
%
    \begin{table*}
    \caption{Mean uncertainty on the DM measurements for each method and pair of ($A_{RN}$, $A_{DM}$) values in the case of $\gamma_{RN}=3.7$, $\gamma_{DM}=8/3$.}
    \centering
    \begin{tabular}{c|ccccc|c}
    \toprule
    ~                 &              &               &      $\langle \sigma_{DM} \rangle\cdot10^{-5} ~~ (\mathrm{pc}\,\mathrm{cm^{-3}})$       &          &        &  \\
    \midrule
      ~     $A_{RN}$            & \textbf{EW}                & \textbf{DMX}               & \textbf{DM GP}             & \textbf{DMX no RN}         & \textbf{DM GP no RN}       & $A_{DM}$ \\ \hline
        $-12.6$ & 5.0 & 1.3  & 1.0 & 0.8 & 0.6 & $-12.3$ \\ 
        $-13.6$ & 5.0 & 1.1  & 0.7 & 0.8 & 0.5 & $-12.3$ \\ 
        $-14.6$ & 5.0 & 1.0  & 0.6 & 0.8 & 0.5 & $-12.3$ \\
        \hline
        $-12.6$ & 5.0 & 1.3  & 1.0 & 0.8 & 0.6 & $-13.3$ \\ 
        $-13.6$ & 5.0 & 1.1  & 0.6 & 0.8 & 0.5 & $-13.3$ \\ 
        $-14.6$ & 5.0 & 1.0  & 0.7 & 0.8 & 0.5 & $-13.3$ \\
        \hline
        $-12.6$ & 5.0 & 1.3  & 0.8 & 0.8 & 0.4 & $-14.3$ \\ 
        $-13.6$ & 5.0 & 1.1  & 0.5 & 0.8 & 0.4 & $-14.3$ \\ 
        $-14.6$ & 5.0 & 1.1  & 0.4 & 0.8 & 0.4 & $-14.3$ \\
        \bottomrule
    \end{tabular}
    \label{tab: errors RN -11/3, DM -8/3}
    \end{table*}
    \begin{table*}
    \caption{Mean uncertainty on the DM measurements for each method and pair of ($A_{RN}$, $A_{DM}$) in the case of $\gamma_{RN}=2.5$, $\gamma_{DM}=3.2$.}
    \centering
    \begin{tabular}{c|ccccc|c}
    \toprule
    ~                 &              &               &      $\langle \sigma_{DM} \rangle \cdot10^{-5} ~~ (\mathrm{pc}\,\mathrm{cm^{-3}}) $      &          &        &  \\
    \midrule
      ~      $A_{RN}$   & \textbf{EW}                & \textbf{DMX}               & \textbf{DM GP}             & \textbf{DMX no RN}         & \textbf{DM GP no RN}       & $A_{DM}$ \\ \hline
        $-12.6$ & 5.0 & 1.3 & 0.9 & 0.8 & 0.6 & -12.3 \\ 
        $-13.6$ & 5.0 & 1.1 & 0.6 & 0.8 & 0.5 & -12.3 \\ 
        $-14.6$ & 5.0 & 1.0 & 0.6 & 0.8 & 0.5 & -12.3 \\
        \hline
        $-12.6$ & 5.0 & 1.3 & 0.9 & 0.8 & 0.5 & -13.3 \\ 
        $-13.6$ & 5.0 & 1.0 & 0.6 & 0.8 & 0.5 & -13.3 \\ 
        $-14.6$ & 5.0 & 1.0 & 0.6 & 0.8 & 0.5 & -13.3 \\
        \hline
        $-12.6$ & 5.0 & 1.3 & 0.7 & 0.8 & 0.4 & -14.3 \\ 
        $-13.6$ & 5.0 & 1.0 & 0.4 & 0.8 & 0.4 & -14.3 \\ 
        $-14.6$ & 5.0 & 1.0 & 0.4 & 0.8 & 0.4 & -14.3 \\ 
        \bottomrule
    \end{tabular}
    \label{tab: errors RN -2.5, DM -3.2}
    \end{table*}

    We present also our results on the achromatic noise recovery in Appendix \ref{Appendix: B}.

\section{Discussion}\label{sec:4}

\subsection{Absorption of red noise in the DM reconstruction}
\begin{figure*}
        \centering
        \includegraphics[width=0.92\textwidth]{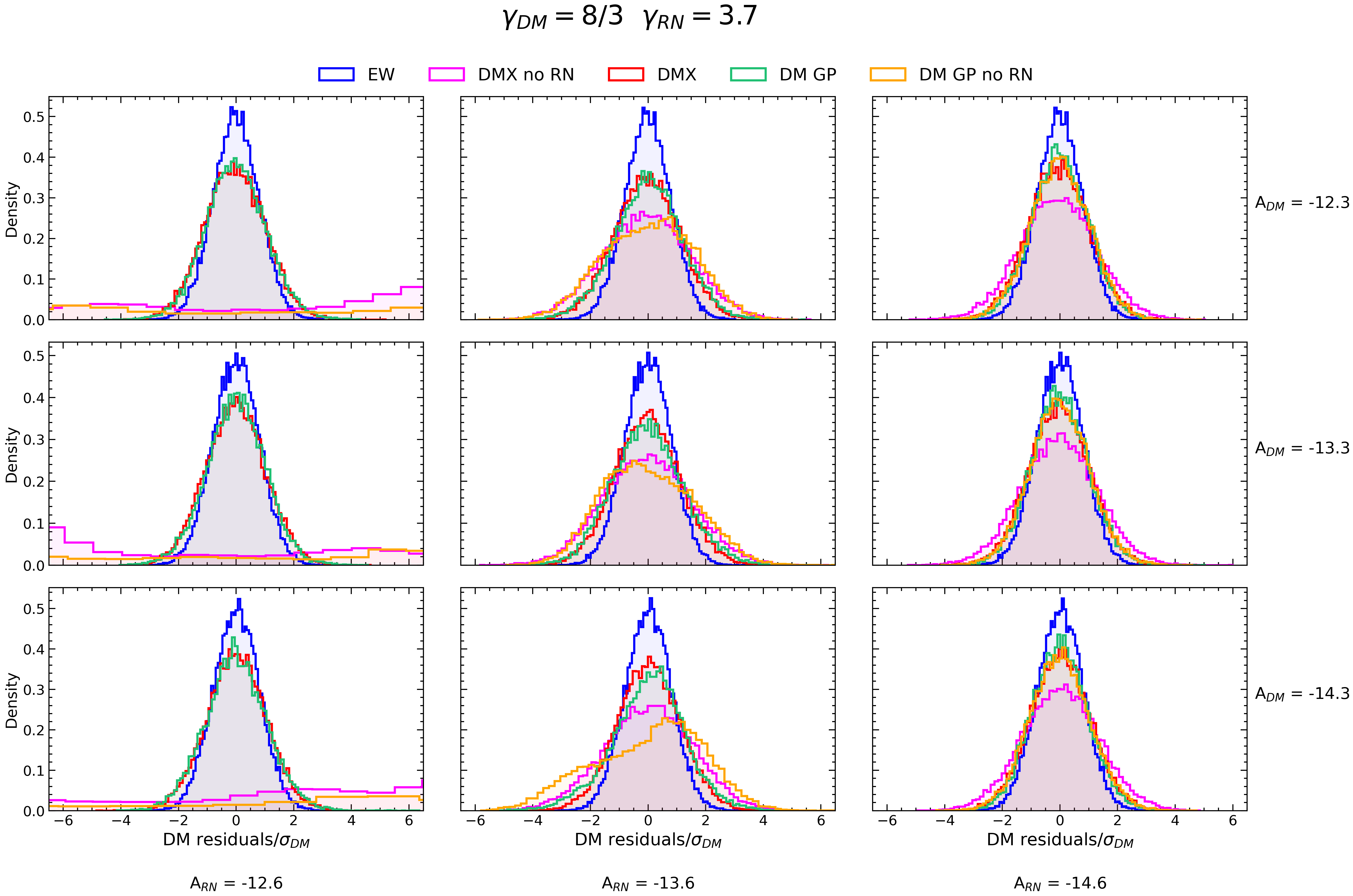}
        \caption{Histograms of DM residuals normalized by their uncertainty for \textbf{EW} (blue), \textbf{DMX no RN} (magenta), \textbf{DM GP no RN} (orange), \textbf{DMX} (red),\textbf{DM GP} (green).
        $A_{RN}$ decreases going left to right (as indicated at the bottom of the columns); $A_{DM}$ increases going upwards (as indicated on the right of each row). Spectral indices are $\gamma_{RN} = 3.7$, $\gamma_{DM}=8/3$.}
        \label{fig: DM residuals hist gamma 3.6}
    \end{figure*}

\begin{figure*}
        \centering
        \includegraphics[width=0.92\textwidth]{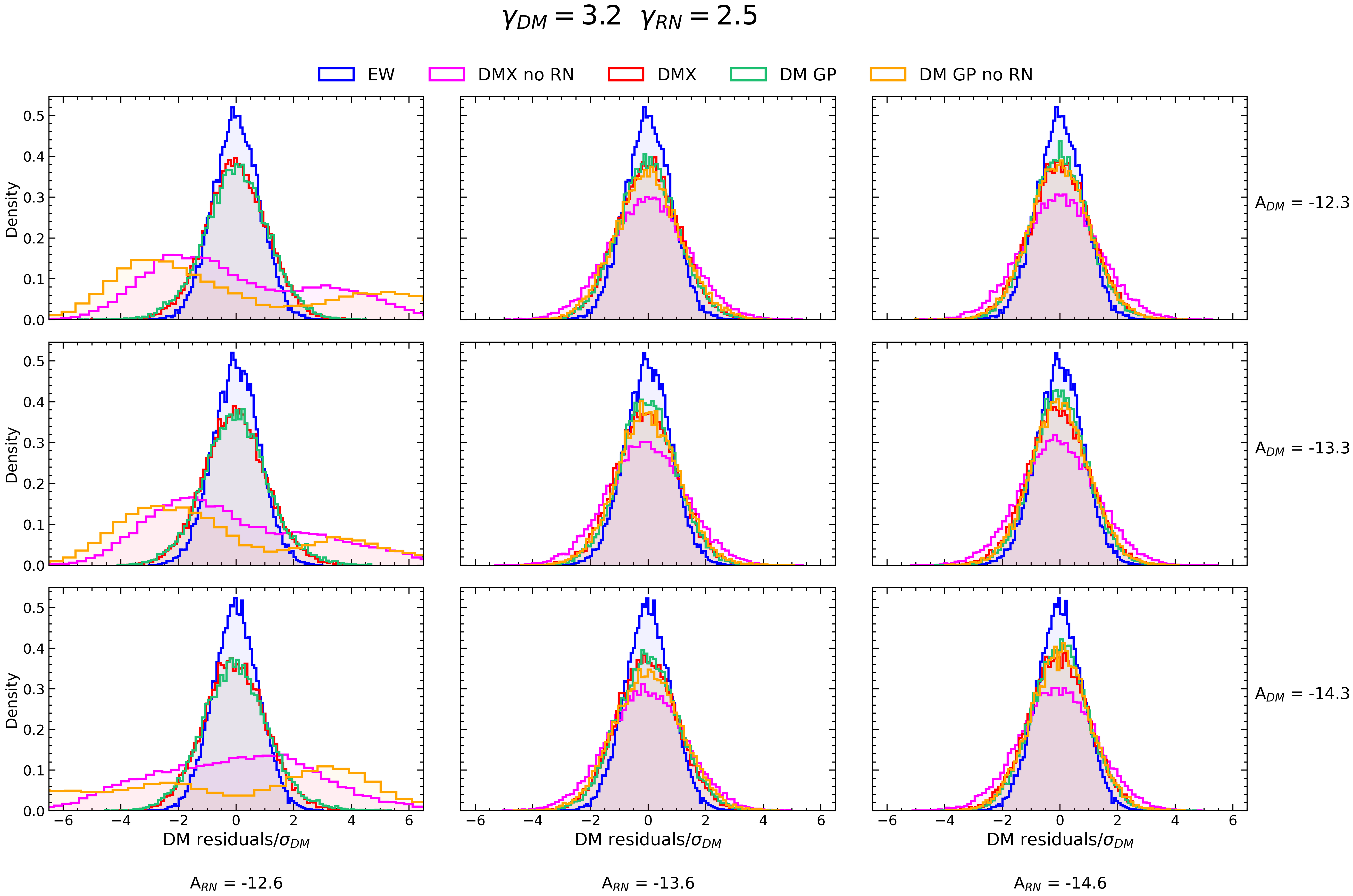}
        \caption{As in Figure~\ref{fig: DM residuals hist gamma 3.6}. Spectral indices are $\gamma_{RN} = 2.5$, $\gamma_{DM}=3.2$.}
        \label{fig: DM residuals hist gamma 2.5}
    \end{figure*}

    The presence of both RN and DM noise in our initial dataset allows us to compare the ability of each method to correctly identify the chromatic noise processes when also achromatic signals are present. 
    The DM residuals reported in the top panel of Figure~\ref{fig: DM residuals 100} show that each method that models RN does not display any evident structure.
    On the other hand, \textbf{DM GP no RN} and \textbf{DMX no RN} show evident corruptions in the DM recovery, whose structure matches the injected RN signal (bottom panel). The same conclusion can be drawn by looking at the first column of Figures~\ref{fig: DM residuals hist gamma 3.6} and \ref{fig: DM residuals hist gamma 2.5}, in which the distributions of the normalized DM residuals calculated for those two methods are not symmetric and not centered around zero.
    This indicates that modeling by \textbf{DM GP} and \textbf{DMX} is not inherently restricted to modeling frequency-dependent signals, but is susceptible to absorbing frequency-independent signals as well, especially if those are not being independently modeled in the analysis. 
    This is true as long as the RN signal is sufficiently `loud' to rise above the WN level.
    When this condition is not satisfied, or when the power spectrum of the RN exceeds the WN power in by only a few frequency bins, then the DM recovery is not as much affected by the absence of RN modeling.
    This can be easily seen in the second and third column of Figures~\ref{fig: DM residuals hist gamma 3.6} and \ref{fig: DM residuals hist gamma 2.5}, where the distributions are very similar to each other because of the low amplitude of the RN ($-13.3$ and $-14.3$ respectively).

\subsection{Comparison of the methods based on precision and accuracy}

    We base the comparison among the methods on both the yielded precision and accuracy.
    We define precision as the uncertainty on the measured DM.
    The method which gives measurements with the lowest uncertainties and hence is the most precise is \textbf{DM GP}, followed by \textbf{DMX} and then \textbf{EW}. 
    This can be inferred from Tables~\ref{tab: errors RN -11/3, DM -8/3} and \ref{tab: errors RN -2.5, DM -3.2}, where we report the mean uncertainty on the DM recoveries per each method and set of parameters. 
    The mean uncertainties of \textbf{DM GP} are around one order of magnitude smaller than the \textbf{EW} ones, while \textbf{DMX} sits in the middle of the two.
    When the noise model lacks the RN component (\textbf{DMX no RN}, \textbf{DM GP no RN}) and hence we fit the residuals only with the DM and WN, the mean DM uncertainty is systematically lower than for the full noise model which includes WN, RN, DM.
    This is expected because without fitting for the RN we are reducing the number of parameters in the model.
    The behavior of the DM residuals when we consider all the 100 simulated realization together is shown in Figure~\ref{fig: DM residuals 100} for all of the five methods.
    The least precise method, \textbf{EW}, is also associated to the largest spread in the residuals, while \textbf{DM GP} (with RN modeled) produces a much smaller deviation of the residuals around zero.
    
    While precision informs about the uncertainty level, it does not report how far the data points are from the true value.
    We define accuracy as the level of agreement between a test result and the true value\footnote{ISO 5725-1:2023 in International Organization for Standardization (\url{https://www.iso.org/home.html})}.
    The accuracy test is conducted by studying the histograms in Figures~\ref{fig: DM residuals hist gamma 3.6} and \ref{fig: DM residuals hist gamma 2.5}.
    These are obtained by normalizing the DM residuals by their uncertainties, hence offering a measure that serves as a proxy for the deviation of each bin in the histogram from the true value in terms of sigmas.

    At first, we check the fraction of residuals that lies within the $3\sigma$ threshold for each method, which for a Gaussian distribution is expected to be $99.7\%$.
    For \textbf{EW} more than $99.9\%$ of the values fall within $3\sigma$, and this happens mostly because the uncertainties on the DM measurements are the highest among the methods.
    \textbf{DMX} is the second most accurate method, with a percentage of data points within 3$\sigma$ greater than $99.5\%$, followed by \textbf{DM GP} with $98.5\%$.
    The remaining methods converge to higher percentages, comparable with the three just mentioned, only when the RN amplitude is at the lowest simulated level ($-14.3$).

\subsection{Whitening of the DM residuals}\label{sec: whitening}
\begin{figure*}
        \includegraphics[width=0.93\linewidth]{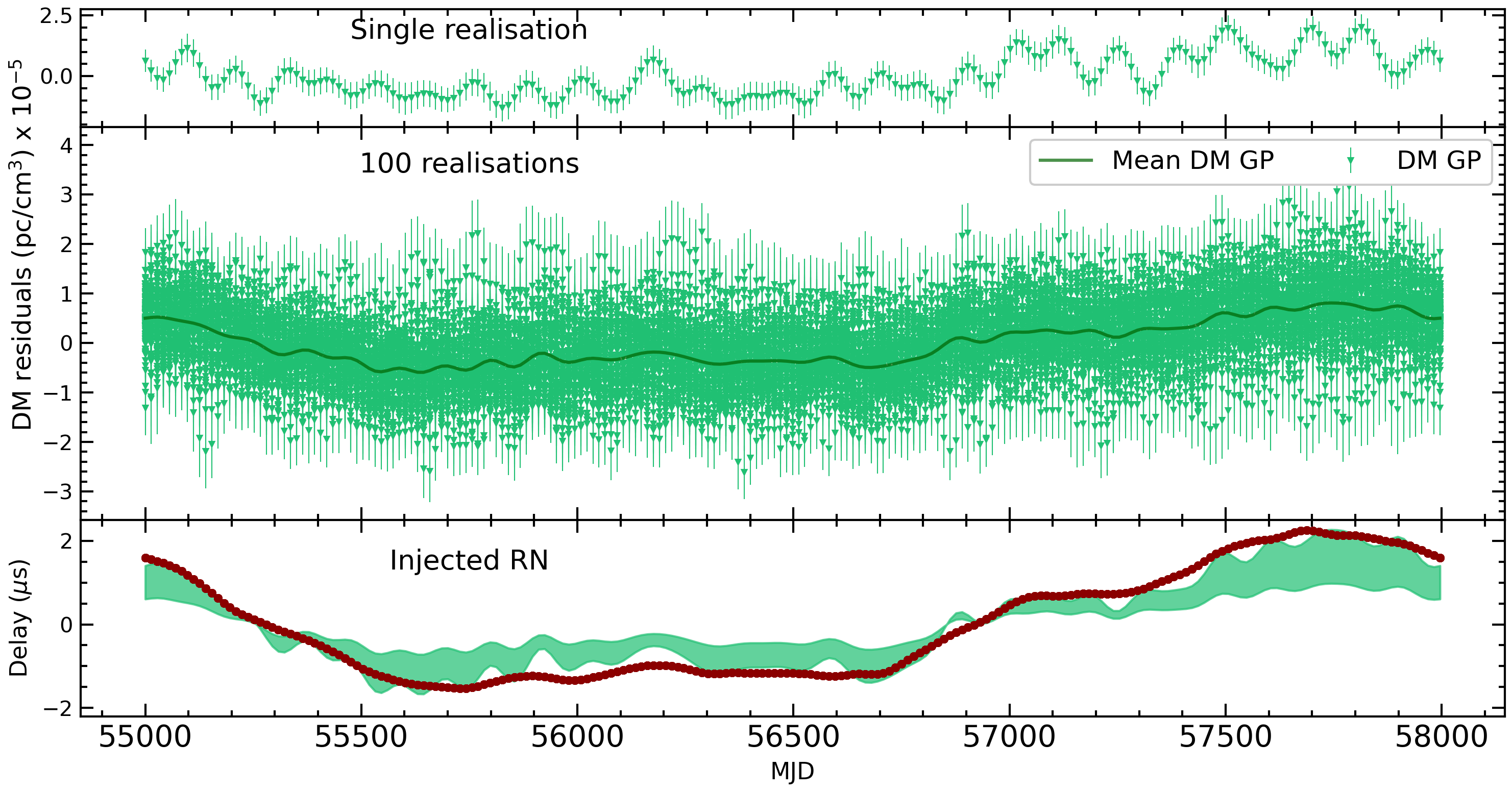}
        \caption{The top panel shows the DM residuals obtained by applying \textbf{DM GP} (with RN modeling) to an individual simulation over the 100 realized for values of the noise parameters $A_{RN} = -13.6$, $A_{DM}=-13.3$, $\gamma_{RN} = 3.7$, $\gamma_{DM}= 8/3$, EFAC and EQUAD as Table~\ref{tab: Parameters} and $\sigma_{\mathrm{ToA}}~=5\mathrm{\mu s}$. 
        The second panel reports an average of the DM residuals left by \textbf{DM GP} over the 100 simulations performed with the same parameters. The bottom panel shows the injected RN signal (red dots) compared to the time delay associated with the mean \textbf{DM GP} (shaded green).
        Note that we have a shaded region because a single DM value causes a frequency-dependent delay.
        Hence, the edges of the green region correspond to the maximum and minimum frequencies of the band.
        }
        \label{fig: GP with RN absorption}
    \end{figure*}

\begin{figure*}
        \centering
        \includegraphics[width=\textwidth]{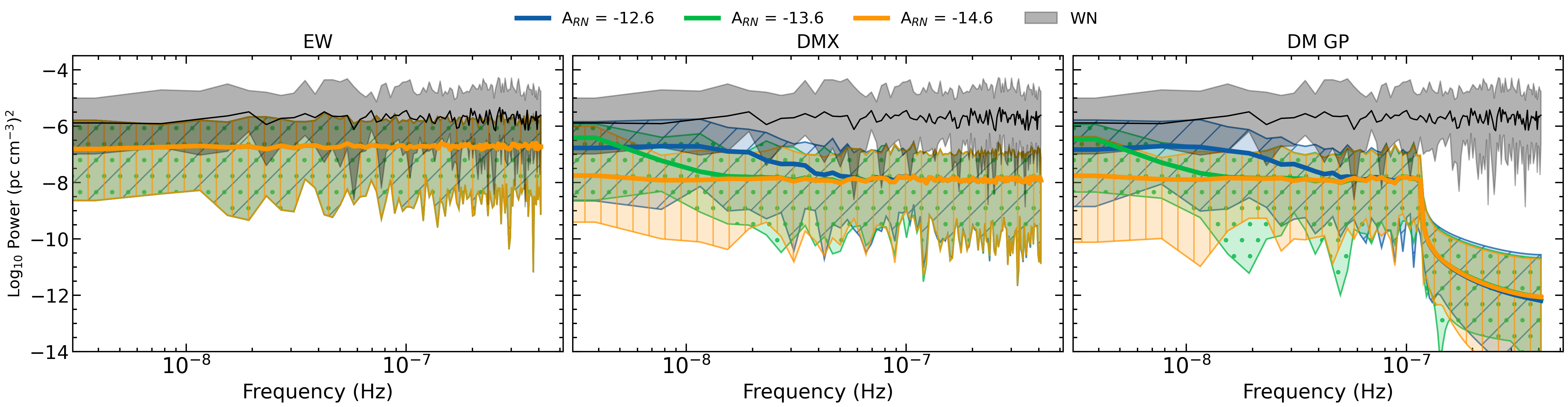}
        \caption{Power spectra of the DM residuals left by the \textbf{EW} (left panel), \textbf{DMX} (central panel), and \textbf{DM GP} (right panel) methods.
        We also show the WN level (black) and the power spectra for different RN amplitudes: $A_{RN}=-14.6$ orange with vertical dashes; $A_{RN}=-13.6$ green with dots; $A_{RN}=-12.6$ blue with oblique dashes.
        With \textbf{DM GP} we model the noise up to a Fourier-frequency of $30/T_{span}\sim1.15\times10^{-7}~\mathrm{Hz}$, the same as the injected signal. 
        That is why in the right-hand panel there is a drop at that specific frequency of the power spectrum of the \textbf{DM GP} residuals.
        }
        \label{fig: fft}
    \end{figure*}

    Other than precision and accuracy, we also investigate the ability of each method to whiten the DM residuals, meaning that we check if they are random independent without correlated structures.
    This is extremely important for noise modeling, since any structure left in the DM residuals is associated with either the absorption of power in other noise components or with artifacts introduced by the method itself.
    We have conducted a Gaussianity test on the normalized DM residuals, and the associated reduced chi-squared values $\chi_{red}^2$ are reported in Tables~\ref{tab: chi2 RN -11/3, DM -8/3} and \ref{tab: chi2 RN -2.5, DM -3.2}.
    For \textbf{EW}, $0.87 \leq \chi^2_{red} \leq 1.35$, which implies that the data are well described by a Gaussian distribution and that there are no, or very few, remaining time-correlated structures in the DM residuals.
    For \textbf{DMX}, $0.71 \leq \chi^2_{red} \leq 1.49$, showing that \textbf{EW} does not outperforms appreciably \textbf{DMX} in whitening the DM residuals.
    For \textbf{DMX no RN} and \textbf{DM GP no RN} the results are as expected:
    as long as $A_{RN}$ is high, they are not able to properly recover the DM values and hence the $\chi^2_{red}$ associated with the histograms of the normalized residuals is very high.
    If $A_{RN} = -13.6$, \textbf{DMX no RN} can reach $\chi_{red}^2$ values which are closer to 1 with respect to \textbf{DM GP no RN}, which does so only when $A_{RN}=-14.6$.
    Lastly, \textbf{DM GP}, even though it is the most precise method, is not the most accurate and also not the best in whitening the DM residuals.
    The distributions in Figures~\ref{fig: DM residuals hist gamma 3.6} and \ref{fig: DM residuals hist gamma 2.5} are most of the times similar to the \textbf{DMX} ones, but the corresponding $\chi_{red}^2$ are not as close to 1.
    \begin{table*}
    \caption{Reduced $\chi^2$ of a Gaussian fit to the histograms reported in Figure~\ref{fig: DM residuals hist gamma 3.6}.  $\gamma_{RN} = 3.7$, $\gamma_{DM}=8/3$.}
    \centering
    \begin{tabular}{c|rrrrr|c}
    \toprule
        &    &      &     $\chi^2_{red}$     &       &                 &            \\
        \midrule
      ~     $A_{RN}$      & \textbf{EW}   & \textbf{DMX}  & \textbf{DM GP} & \textbf{DMX no RN} & \textbf{DM GP no RN} & $A_{DM}$ \\ 
      \hline
        $-12.6$    & 0.95 & 1.37 & 1.53  & 1108.49   & 2691.50     & -12.3    \\ 
        $-13.6$    & 0.91 & 1.04 & 1.73  & 1.34      & 18.75       & -12.3 \\ 
        $-14.6$    & 0.91 & 1.10 & 1.21  & 0.80      & 1.25        & -12.3 \\
        \hline
        $-12.6$    & 0.87 & 0.78 & 1.67 & 6154.89    & 6538.24     & -13.3 \\ 
        $-13.6$    & 0.88 & 1.01 & 2.45 & 1.33       & 26.07       & -13.3 \\ 
        $-14.6$    & 0.90 & 0.84 & 1.35 & 0.76       & 1.00        & -13.3 \\
        \hline
        $-12.6$    & 0.89 & 0.86 & 1.63 & 1104.90      & 1579.96   & -14.3 \\ 
        $-13.6$    & 0.89 & 0.71 & 5.25 & 1.53         & 71.90     & -14.3 \\ 
        $-14.6$    & 0.87 & 0.83 & 1.26 & 0.73         & 0.93      & -14.3 \\
        \bottomrule
    \end{tabular}
    \label{tab: chi2 RN -11/3, DM -8/3}
    \end{table*}
    \begin{table*}
    \caption{Reduced $\chi^2$ of obtained from a Gaussian fit on the histograms reported in Figure~\ref{fig: DM residuals hist gamma 2.5}.   $\gamma_{RN} = 2.5$, $\gamma_{DM}=3.2$.}
    \centering
    \begin{tabular}{c|rrrrr|c}
    \toprule
        &    &      &     $\chi^2_{red}$     &       &                 &            \\
        \midrule
      ~   $A_{RN}$                & \textbf{EW}   & \textbf{DMX}  & \textbf{DM GP} & \textbf{DMX no RN} & \textbf{DM GP no RN} & $A_{DM}$ \\ \hline
        $-12.6$    & 1.35 & 0.98 & 1.96  & 365.57    & 218.71      & -12.3    \\ 
        $-13.6$    & 1.35 & 0.82 & 1.45  & 0.90      & 1.99        & -12.3 \\ 
        $-14.6$    & 1.35 & 1.03 & 1.09 & 0.94       & 0.98        & -12.3 \\
        \hline
        $-12.6$    & 1.17 & 1.49 & 3.35 & 168.97     & 194.42      & -13.3 \\ 
        $-13.6$    & 1.16 & 1.32 & 1.58 & 1.02       & 1.20        & -13.3 \\ 
        $-14.6$    & 1.17 & 0.76 & 1.13 & 0.97       & 1.12        & -13.3 \\
        \hline
        $-12.6$    & 0.98 & 1.29 & 2.20 & 33.12      & 141.30      & -14.3 \\ 
        $-13.6$    & 0.98 & 0.96 & 0.95 & 0.74       & 1.79        & -14.3 \\ 
        $-14.6$    & 0.98 & 0.95 & 0.77 & 0.77       & 1.45        & -14.3 \\
        \bottomrule
    \end{tabular}
    \label{tab: chi2 RN -2.5, DM -3.2}
    \end{table*}
    Time correlated structures, both short and long-term, are still present in \textbf{DM GP} DM residuals, emphasized by the aforementioned reduced errorbars.
    The short-term time correlated structures are caused by the high Fourier-frequency components of the injected signal (the highest of which is $f_{30}=30/T_{\text{span}}$), while long-term ones are mainly due to the presence of RN in the data.
    A more explicit representation of this feature is shown in Figure~\ref{fig: GP with RN absorption}, 
    where the first two rows show DM residuals for a single and 100 realizations, respectively, using \textbf{DM GP} (similar results to \textbf{DM GP} can be obtained with \textbf{DMX}). 
    The third row reports the comparison between the structure left in the DM residuals and the injected RN, showing that the DM modeling is indeed absorbing power from the RN process, as it displays the same signature of the injected RN signal.
    However, we also need to consider whether these structures have a significant impact on the timing residuals by comparing their power spectrum with the WN level.
    To do so we computed a discrete Fourier transform and subsequently the power spectrum of the DM residuals at different $A_{RN}$ and compared it with the level of the WN injected for each of the three methods.
    The result is shown in Figure~\ref{fig: fft}.
    On average, the WN dominates over the DM residuals which confirms that signature remaining in the DM residuals cannot significantly be detected, and therefore it will not affect the timing residuals.
    On the other hand, we can say that \textbf{EW} is not affected at all by the strength of the RN, while it significantly influences \textbf{DM GP} and \textbf{DMX}.

    Importantly, Figure~\ref{fig: fft} shows that the remaining power associated with the DM residuals is the lowest for \textbf{DM GP} (and \textbf{DMX}). 
    This hints that \textbf{DM GP} (and \textbf{DMX}) diminishes the the dispersive noise level in the ToAs to the minimum among the techniques tested, a result that would be beneficial in studies that need the most optimal noise mitigation possible.
    Hence, while a full study of the \textbf{DM GP} analysis and its interaction with various types of RN (particularly GWs)  is beyond the scope of this paper, this finding indicates that \textbf{DM GP} (and \textbf{DMX}) might work best for whitening timing residuals, for example in a context such as the search for low-frequency GWs.


    \subsection{Alternative simulation settings}

    Here we tested the methods on datasets that we simulated by varying specific settings: first, we reduced the uncertainties of the simulated ToAs from our initial value (from $5\mu\mathrm{s}$ to $1\mu\mathrm{s}$ and $0.1 \mu\mathrm{s}$) to improve our sensitivity to the RN process and DM variations, then we simulated the ToAs at L-band frequencies and finally we changed the injected DM time series.

    \subsubsection{Testing different ToA uncertainties}

    We used $\sigma_{\text{temp}}=5\mu\text{s}$ since it represents the typical mean uncertainties on ToAs obtained from LOFAR observations \citep{Donner2020}. 
    However, there might be pulsars with higher ToA precision, and we expect that future facilities such as the low-frequency part of the Square Kilometer Array \citep{Janssen2015} will improve the ToA sensitivity.
    For this reason, we run two new simulation sets, with decreased values for the $\sigma_{\text{temp}}$, firstly at $1\mu s$ and then at $0.1\mu s$.
    Furthermore, as this reduces the WN floor level, we enhance the methods' capability to identify RN and properly disentangle it from DM variations.
    The WN parameters injected in the simulations were also changed to: $EFAC = 1$, $EQUAD=10^{-8} \,s$.
    Chromatic and achromatic red processes were set both to 30 Fourier-frequency components, with $\gamma_{RN} = 3.7$, $\gamma_{DM} = 8/3$, $A_{RN} = -13.6$ and $A_{DM} = -13.3$.
     
    In Table~\ref{tab: 1 and 0.1 us errors} we report the DM uncertainty levels for each method and each uncertainty.
    Given the smaller errorbars of the ToAs, the precision on the DM measurements increased, as expected.
    As we did previously, we can rank \textbf{DM GP} as the most precise method followed by \textbf{DMX} and \textbf{EW}.
   
    We recall that an important finding from our simulations (see Section~\ref{sec: whitening}) was about the RN signal left in the DM residuals. In fact, Figure~\ref{fig: GP with RN absorption} (bottom panel) shows that the signal left in the DM residuals matches exactly the one of the injected RN. The corresponding delay, that is thus left on average in the timing residuals, was about $\pm2 ~\mu\mathrm{s}$.
    This changes in the two new cases characterized by low ToA uncertainties:
    when $\sigma_{\text{temp}} = 1~\mu\mathrm{s}$ the mean DM signal that is left due to an erroneous RN absorption contributes to the timing residuals at the level of $\pm 150~\mathrm{ns}$; when $\sigma_{\text{temp}} = 0.1~\mu\mathrm{s}$ this becomes $\pm 20~\mathrm{ns} $.
    Since the injected RN signal was the same, we can confirm that higher precision on the ToAs allows to increase the sensitivity to the RN and improve the accuracy of the measurements of DM variations.
    \begin{table}
    \caption{Mean DM uncertainty measured with the \textbf{EW}, \textbf{DMX} and \textbf{DM GP} methods over simulated ToAs with uncertainties fixed at $1\mu $s and $0.1 \mu $s}
        \begin{tabular}{c|ccc}
        \toprule
          $\sigma_{\text{temp}}$ &   &   $\langle \sigma_{DM} \rangle\cdot10^{-6} \quad (\mathrm{pc}\,\mathrm{cm^{-3}})$  & \\
        \midrule
           & \textbf{EW} & \textbf{DMX}  & \textbf{DM GP} \\
         $1\,\mu $s  & $6.3$ & $2.0$ & $1.6$ \\
        $0.1\,\mu $s  & $0.6$ & $0.5$ & $0.2$ \\
        \bottomrule
        \end{tabular}
        \label{tab: 1 and 0.1 us errors}
    \end{table}

    \subsubsection{Testing ToAs at L-band}
    \label{subsec: L-band}  
    In this part we report the results obtained by applying the DM-computation methods to L-band ToAs.
    To simulate high-frequency ToAs, we refer to the data collected by the Nan\c{c}ay decimeter radiotelescope (NRT), which has a relatively large bandwidth (512~MHz) centered at 1.4~GHz divided in 4 subbands \citep{EPTA2023paperI}.
    The input parameter are: EFAC $=1$, $\text{EQUAD}=10^{-8} \text{s}$, $A_{DM} = -13.3$ and $\gamma_{DM} = 8/3$.
    We use a conservative $1~\mu\mathrm{s}$ ToA template-fitting error.
    
    Concerning the precision of the methods, the results are similar as for the LOFAR frequency case: \textbf{DM GP} is the most precise, followed by \textbf{DMX} and \textbf{EW}.
    However, the size of these errors is now larger by approximately two orders of magnitude (see Table~\ref{tab: Nancay errors}). 
    This is due to two main reasons: the first is the increased central frequency of the analysed ToAs (which is $1.4~\mathrm{GHz}$ against the $153~\mathrm{MHz}$ of the previous LOFAR-like case), because DM is inversely proportional to the squared of the observing frequency.
    The second is the fractional bandwidth $B$ (see Equation~\ref{eq: js18}):
    the higher $B$, the better the precision in the DM measurements (\citealt{vs18}). 
    For the NRT-like case $B \simeq 0.34$, while for the LOFAR-like case $B \simeq 0.41$.
    \begin{table}
    \caption{As in Table~\ref{tab: 1 and 0.1 us errors} but for L-band ToAs with $\sigma_{\text{temp}}=1~\mu\mathrm{s}$ and injected DM noise parameters: $A_{DM}=-13.6$, $\gamma_{DM}=3.7$.}
    \centering
        \begin{tabular}{ccc}
        \toprule
              &   $\langle \sigma_{DM} \rangle\cdot10^{-4} \quad (\mathrm{pc}\,\mathrm{cm^{-3}})$  & \\
        \midrule
            \textbf{EW} & \textbf{DMX}  & \textbf{DM GP} \\
            $7.8$ & $2.1$ & $0.7$ \\
        \bottomrule
        \end{tabular}
        \label{tab: Nancay errors}
    \end{table}
    Concerning accuracy, we also report similar conclusions as before: \textbf{EW} has more than $99.76\%$ of the values within $3\sigma$ and it is the most accurate method, followed by \textbf{DMX} with more than $99.44\%$ and finally \textbf{DM GP} with $99.2\%$.
    As well, the Gaussianity test provides similar results to the LOFAR-like case. 
    In Table~\ref{tab: Nan\c{c}ay chi2} are listed the reduced $\chi^2$ values which confirm that on average \textbf{EW} and \textbf{DMX} are working marginally better in whitening the residuals due to their larger errorbars (and thus lower precision), on single DM recoveries.
    \begin{table}
    \caption{Reduced $\chi^2$ of the normalized DM residuals.}
        \centering
        \begin{tabular}{ccc}
        \toprule
              &   $\chi^2_{red}$  & \\
        \midrule
            \textbf{EW} & \textbf{DMX}  & \textbf{DM GP} \\
            $1.07$ & $1.10$ & $1.54$ \\
        \bottomrule
        \end{tabular}
        \tablefoot{It has been obtained with a Gaussian fit on histograms of DM residuals normalized by their uncertainty for the three main tested methods (\textbf{EW}, \textbf{DMX} and \textbf{DM GP}) over ToAs simulated at L-band with $\sigma_{\text{temp}}=1~\mathrm{\mu s}$ and injected DM noise with $A_{DM}$, $\gamma_{DM}$ of, respectively, $-13.3$ and $8/3$.}
        \label{tab: Nan\c{c}ay chi2}
    \end{table}
    In the top panel of Figure~\ref{fig: Nan\c{c}ay res_and_hist} are presented the DM residuals for each of the three methods, while the corresponding normalized histograms are reported in the bottom plot.
    \begin{figure}
        \centering
        \includegraphics[width=\linewidth]{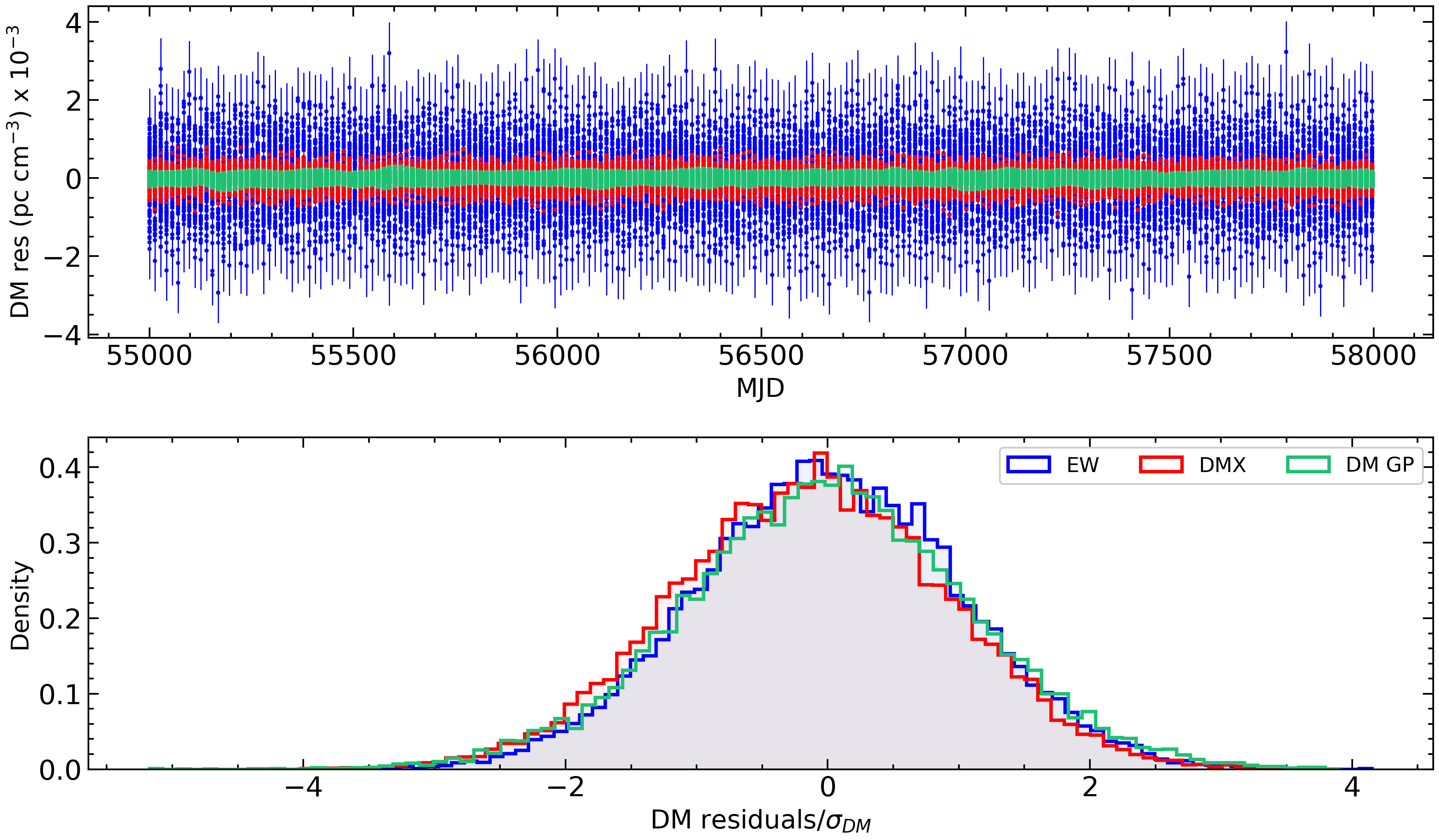}
        \caption{Top panel: DM residuals left by the \textbf{EW} (blue), \textbf{DMX} (red), \textbf{DM GP} (green) methods when run over the 100 realizations of L-band ToAs. Bottom panel: histograms of the DM residuals normalized by their uncertainty. The input noise parameters are:  $A_{DM}=-13.3$, $\gamma_{DM}=8/3$}
        \label{fig: Nan\c{c}ay res_and_hist}
    \end{figure}
    Figure \ref{fig: Nancay FFT} shows the power spectra of the DM residuals compared with the injected signal.
    Despite the fact that the WN is dominating in most of the frequency bins, all of the three methods can model the long-term structures present in the data.
    At the highest frequencies in the spectra, \textbf{EW} and \textbf{DMX} cannot match the injected signal anymore due to the limits imposed by the WN.
    Instead, \textbf{DM GP} is the only method that produces a power spectrum which does not cross the injection line, meaning that it is not adding any more power at those frequencies.
    Note that \textbf{DM GP} uses a power-law model to describe the noise process, which is identical to the functional form of the injected signal.
    Therefore, it is possible that the reported results show a biased result for \textbf{DM GP}, since in real-life datasets we may not have pure power-law noise processes.

    To evaluate a more realistic case, in the following section we test the methods on an injected DM time series which does not come from a pure power-law.

    \begin{figure*}
        \centering
        \includegraphics[width=\textwidth]{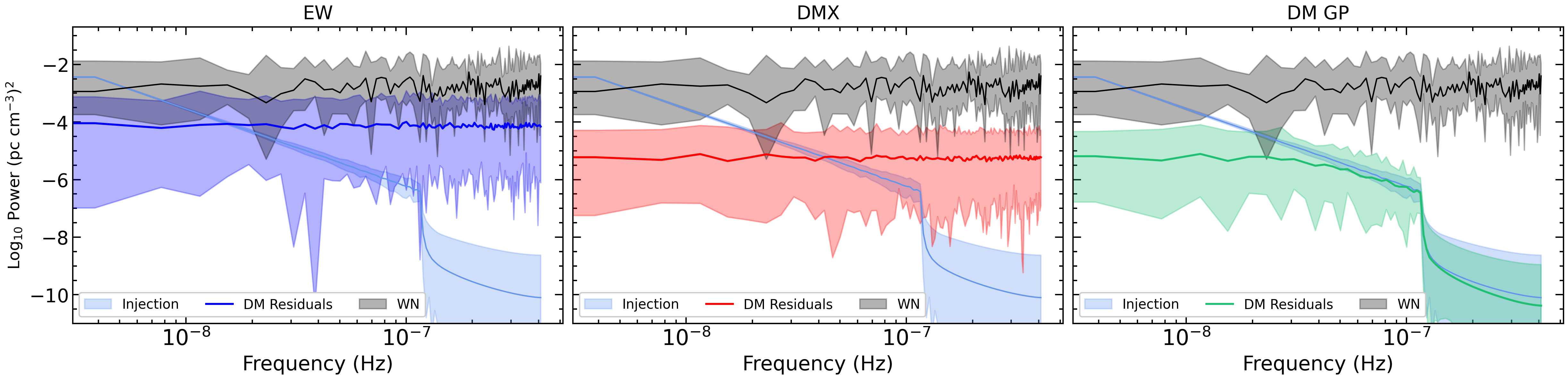}
        \caption{Power spectra of the DM residuals obtained using  \textbf{EW} (left), \textbf{DMX} (center), \textbf{DM GP} (right), compared with the injected signal (light blue) at L-band. Input noise parameters: $A_{DM}=-13.3$, $\gamma_{DM}=8/3$.
        The injected signal power (light-blue) drops at a frequency of $30/T_{span}~\sim~1.15~\times~10^{-7}~\mathrm{Hz}$.
        We use the same number of frequency components in the modeling with \textbf{DM GP}.}
        \label{fig: Nancay FFT}
    \end{figure*}

    \subsubsection{Custom DM injection}
    \label{subsec: custom DM}

\begin{figure}
        \centering
    \includegraphics[width=\linewidth]{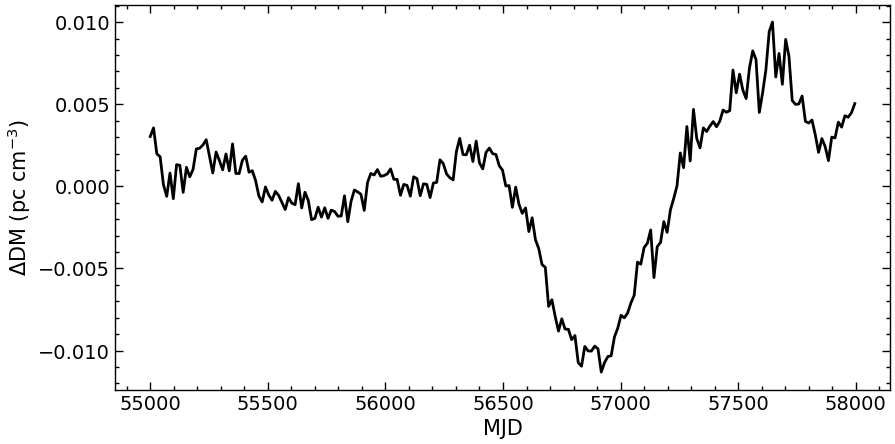}
        \caption{DM time series injected in the data with an empirical power spectrum. \citep{jdthesis_2022}}
        \label{fig: custom DM}
    \end{figure}

\begin{figure}
        \centering
        \includegraphics[width=\linewidth]{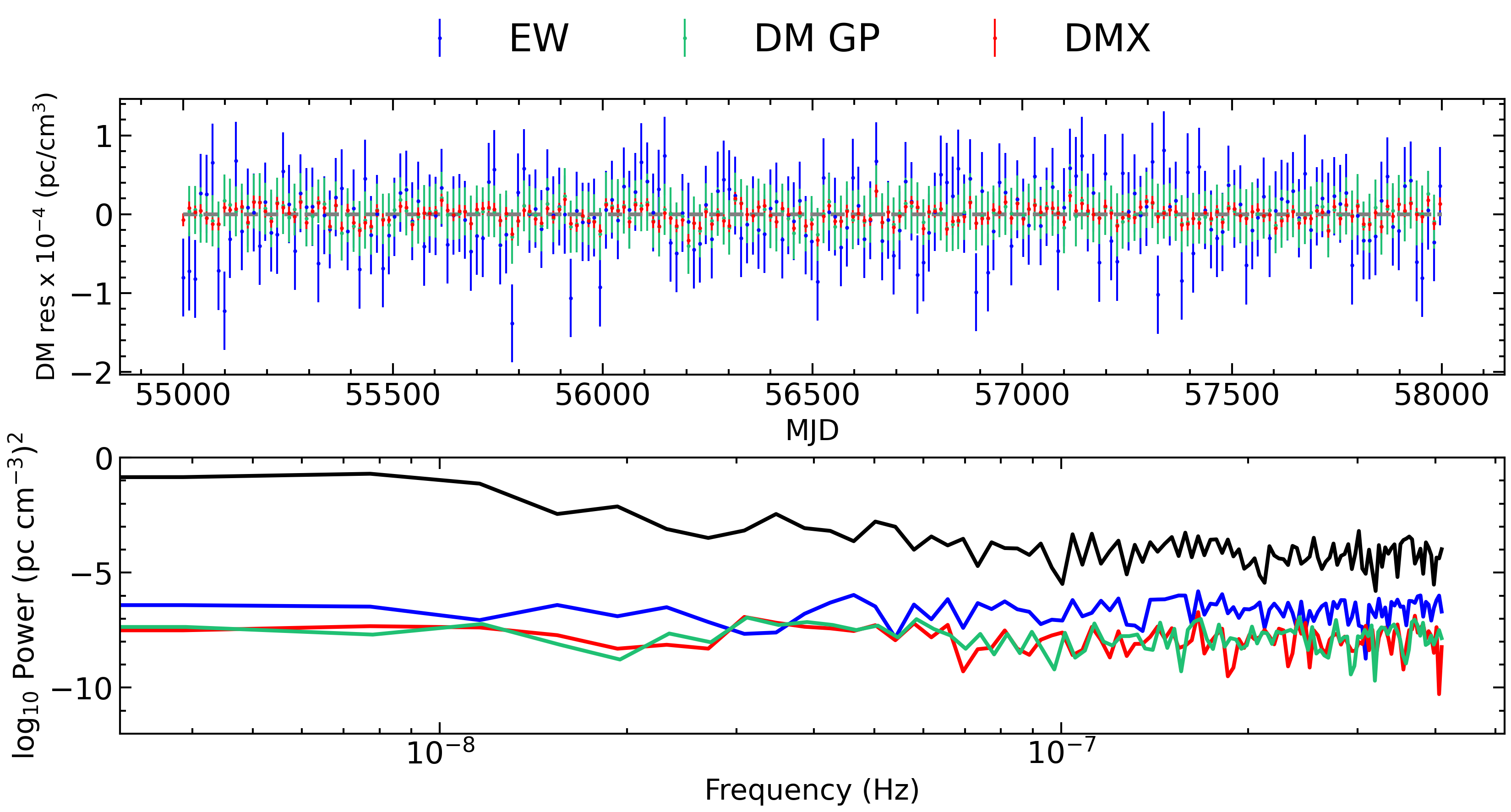}
        \caption{Top panel: DM residuals for \textbf{EW} (blue), \textbf{DM GP} (green), \textbf{DMX} (red) obtained by taking the difference of the DMs recovered by each method and the injected DM time series of Figure~\ref{fig: custom DM} with unknown power spectrum.
        WN parameters are EFAC~$=1.2$ and EQUAD~$2\mu\mathrm{s}$ and ToA uncertainty set at $5\mu\mathrm{s}$.
        Bottom panel: Power spectra of the injected signal (black) and of the DM residuals.}
        \label{fig: residuals and PS custom DM}
    \end{figure}

    Here we test the performance of the three considered techniques over an individual, injected DM time-series with an empirical power-spectrum shape, derived in \cite{jdthesis_2022} from the LOFAR dataset of PSR~J0139+5814 (see Figure~\ref{fig: custom DM}).
    The ToAs are still evenly spaced, with an uncertainty set to $5\mu\text{s}$ and the injected WN is defined by $\text{EFAC}=1.2$ and $\text{EQUAD}=2 \mu\text{s}$.
    Given the short timescale fluctuations in the injected signal, we use 107 Fourier components to define the power-spectrum used in \textbf{DM GP}, that corresponds to the limit imposed by the Nyquist theorem given the 215 data points.
    The result of the analysis are reported in Figure~\ref{fig: residuals and PS custom DM} where we show the DM residuals in the top panel and their power spectra in the bottom panel.
    As previously noted, \textbf{EW} and \textbf{DMX} provide independent DM measurements \citep[although \textbf{DMX} might introduce some degree of correlations for close-by points, see][]{Lam2017}, while \textbf{DM GP} assumes an underlying model for the power spectrum to describe the DM variations. 
    The \textbf{DM GP} errorbars are now larger than the \textbf{DMX} ones and hence the recoveries are extremely accurate.
    The bottom panel of Figure~\ref{fig: residuals and PS custom DM} shows that even with the injected signal resembling a broken power law, \textbf{DM GP} was able to absorb the signal and hence successfully recover the injected DM time series.

\section{Conclusion}\label{sec:5}

    In this paper, we carried out a series of simulations in order to test the performance of the three main methods available in the PTA community to calculate DM variations: \textbf{EW}, \textbf{DM GP} and \textbf{DMX}. 
    The final aim is to provide information on the calculation of the DM time series for, e.g., studies of the IISM and the Solar wind.
    

    %

    Our main conclusion is that, while they all perform well, the most accurate method to calculate DM variations is \textbf{EW}, 
    but \textbf{DM GP} and \textbf{DMX} seem to lower the level of the DM noise in the timing residuals the most.

    \vspace{0.2cm}
    
    In more detail, we conclude that all of the methods perform satisfactorily (as long as the RN and DM models are selected correctly for \textbf{DMX} and \textbf{DM GP}).
    Figure \ref{fig: DM residuals 100} demonstrates that when the RN power is significant, it is unavoidable to include it in the noise model, otherwise the achromatic, time-correlated structures will be absorbed in the DM modeling in the cases of \textbf{DMX} and \textbf{DM GP}.

    With respect to the precision, the mean uncertainties $\langle \sigma_{DM} \rangle$ on the DM recoveries for \textbf{EW} are independent of the amplitudes of the red noise processes.
    This means that the method is only affected by the ToA precision and the frequency coverage of the data.
    For \textbf{DMX} and \textbf{DM GP}, instead, $\langle\sigma_{DM}\rangle$ depends on the amplitude of the injected RN process (see Tables~\ref{tab: errors RN -11/3, DM -8/3} and \ref{tab: errors RN -2.5, DM -3.2}).
    Overall, the method which gives the most precise DM measurements is \textbf{DM GP}.

    We also studied the accuracy, defined as the level of agreement between a test result and the true value while taking into account the uncertainties, by analyzing the distributions of the normalized DM residuals (Figures~\ref{fig: DM residuals hist gamma 3.6} and \ref{fig: DM residuals hist gamma 2.5}).
    Because of the larger errorbars, \textbf{EW} is the most accurate method having in most of the cases $>$99.9\% of the data within 3$\sigma$.
    The second most accurate method, according to this criterion, is \textbf{DMX} and in the end \textbf{DM GP}, which presents more data points in the tail of the distribution because of its high DM precision.
    This is true not only at LOFAR frequencies, but also in the L-band trials, where we simulated a NRT-like frequency coverage.

    The accuracy test, though, as described above, does not imply that the DM residuals are white.
    Hence, we have conducted a Gaussianity test on each of the normalized DM residuals distributions, which informs on the ability of each method in removing time-correlated structures from the ToAs.
    The $\chi_{red}^2$ of the Gaussian fits are reported in Tables~\ref{tab: chi2 RN -11/3, DM -8/3} and \ref{tab: chi2 RN -2.5, DM -3.2}, and they show that \textbf{EW} is the method with the narrowest range of $\chi_{red}^2$ around 1.
    \textbf{DMX} shows similar results, but only when the RN is modeled.
    However, the $\chi_{red}^2$ range shown by \textbf{DMX} fluctuates more than for \textbf{EW}.
    \textbf{DM GP}, which uses a linear combination of Fourier components, produces a time series where close data points are not completely independent. This affects the overall residual distribution, resulting in the widest $\chi_{red}^2$ range spanned among the three methods. 
    
    To better understand the reasons behind the \textbf{DMX} and \textbf{DM GP} distributions, we studied the effect of the RN on the DM residuals.
    In Figure~\ref{fig: GP with RN absorption} we show that even when the RN is modeled, a signature can still be found in the DM residuals yielded by \textbf{DM GP} and \textbf{DMX}. This happens when the WN level exceeds the RN, especially at the highest Fourier frequencies.
    Nonetheless, Figure~\ref{fig: fft} shows that, at LOFAR frequencies, the unmodeled RN signal remains below the WN level.
    In order to properly disentangle the RN and DM noise signal it is necessary to have high-precision ToAs. In fact, we investigated this scenario by simulating datasets with reduced uncertainties,  $\sigma_{ToAs}=1~\mu s, \, 0.1~\mu s$, and, as expected, \textbf{DMX} and \textbf{DM GP} do not show any significant structures related to the injected RN signal in the DM residuals anymore.
    The aforementioned precision-accuracy hierarchy holds in this simulation scenario as well. 
    Similar conclusions are reached in (albeit limited) simulations carried out at L-band and with an injected DM time series with empirical spectral shape.


    \vspace{0.2cm}
    
    However, we stress that these results are based on simulations that do not take into account uneven observational cadence, variable ToA uncertainty, poor, if not absent, frequency coverage of the data and also combination of datasets, i.e., the most common characteristics of real-life data. 
    When these additional complexities are present, it is possible that \textbf{DM GP} performs better than other methods thanks to its flexibility and the fact that its characteristics allow a more optimal blending of data with diverse properties.
    Moreover, the performance of \textbf{EW} and \textbf{DMX} might be affected by more realistic datasets because of their windowing constraints.
    A firmer conclusion will be obtained once these scenarios are taken into consideration.
    Also, the impact of these DM recovery schemes on timing model parameters has not been assessed in our work, but \citealt{Kramer2021} already showed that astrometric parameters can be corrupted by some of the schemes tested. Consequently, future work on this topic would require a complete assessment of that aspect, as well.
    
    Last, we reached the important conclusion that \textbf{DM GP} and \textbf{DMX} seem to have the capacity to reduce the DM noise level to the minimum (see Figure~\ref{fig: fft}). Future works should assess the effectiveness of the analyzed methods in modeling DM variations as a source of noise when a GW background is also present in the data.

\begin{acknowledgements}
FI is supported by the University of Cagliari (IT). FI, CT, AP are supported by the Istituto Nazionale di Astrofisica.
ACh and GMS acknowledge financial support
provided under the European Union’s H2020 ERC Consolidator
Grant “Binary Massive Black Hole Astrophysics” (B Massive, Grant Agreement: 818691).
JPWV acknowledges support from
NSF AccelNet award No. 2114721.
SCS acknowledges the support of a College of Science and Engineering University of Galway Postgraduate Scholarship.
MTL graciously acknowledges support received from NSF AAG award number 2009468, and NSF Physics Frontiers Center award number 2020265, which supports the NANOGrav project.
Work at NRL is supported by NASA.

\end{acknowledgements}

\bibliographystyle{aa} 
\bibliography{./bib}

\begin{appendix}
\section{Evaluating time-domain realizations of stochastic time-correlated signals with LaForge}\label{Appendix LaForge}
This part describes the method used in \textsc{LaForge} software to obtain time-domain realization of stochastic signals measured in the frequency domain. The stochastic and time-correlated signals in PTA data such as DM variations are mainly modeled as Gaussian processes (GPs). GPs are defined as a collection of (infinite) random variables representing the function values (e.g., time delay of the modeled signal in the timing residuals) for all input locations (i.e., observing epochs). Any finite set of these random variables has a joint Gaussian distribution. Thus, GPs allows us to define the mean value and the variance of the signal at any input location. \\
GPs can be fully specified with one of each following approaches \citep{rasmwill2006}:
\begin{itemize}
    \item Weight-Space View, where the process is defined as a linear sum of deterministic basis functions $T_\mu(t)$ multiplied by weights $a_\mu$ as
    \begin{align}\label{eq:fweight}
        f(t) \sim \sum_\mu^m a_\mu \ T_\mu (t),
    \end{align}
    with $\mu = 1, ..., m$ and where the weights are Gaussian random variables as $a_\mu \sim \mathcal{N} \left( a_\mu^0, \phi_{\mu \nu} \right)$, with $a_\mu^0$ assumed to be a zero vector in our case, and $\phi_{\mu \nu}$ corresponding to the weights covariance matrix.\\

    The PTA likelihood including a GP described in the weight-space view can be written as 

    \begin{align} 
    	&p(\delta t | a_\mu, GP) = \nonumber\\
     &\frac{
    		{\rm{exp}} \left[-\frac1 2 \ \sum_{i,j} \
    			\left( \delta t_i - \sum_\mu a_\mu \ T_\mu(t_i) \right) \
    			N_{ij}^{-1} \
    			\left( \delta t_j - \sum_{\mu} a_\mu \ T_\mu(t_j) \right) \right]
    	}
    	{\sqrt{(2\pi)^n \ {\rm{det}}(N)}} \nonumber \\
    	&\times
    	\frac{
    		{\rm{exp}} \left[-\frac1{2} \ \sum_{\mu, \nu} \ a_\mu \ \phi_{\mu \nu}^{-1} \	a_\nu \right]
    	}
    	{\sqrt{(2\pi)^m \ {\rm{det}}(\phi)}},
    	\label{Eq:weightGP}
    \end{align}

    with $\delta t$ and $N$ respectively corresponding to the timing residuals and the covariance matrix, $i, j = 1, ..., n$ and $\mu, \nu = 1, ..., m$.
    
    \item Function-Space View, where the GP directly describes the targeted signal as
    \begin{align}
        f(t) \sim GP \left( g(t), \ k(t, t') \right),
    \end{align}
    including the mean function $g(t)$ and the covariance function $k(t, t')$, also referred to as kernel.\\

    The PTA likelihood in that form is marginalized over the basis weights mentioned above and can be expressed as

    \begin{align}
    	p(\delta t| \text{GP}) = 
    	\frac{
    		{\rm{exp}} \left[-\frac1 2 \ \sum_{ij} \
    			\delta t_i \ (N_{ij} + K_{ij})^{-1} \
    			\delta t_j \right]
    	}
    	{\sqrt{(2\pi)^n  \ {\rm{det}}(N + K)}}, 
    	\label{Eq:MargGP}
    \end{align}    

    where $K = k(t, t')$.
    
\end{itemize}

In PTA analysis, we usually employ Equation \ref{Eq:MargGP} and express $k(t, t')$ from the correspondence between both views as \citep{rasmwill2006}
\begin{align}
    k(t, t') = \sum_{\mu, \nu} \ T_\mu (t) \ \phi_{\mu \nu} \ T_\nu (t'),
\end{align}
In practice, we set the basis functions $T$ as a finite set of sine/cosine functions and 
\begin{align}\label{eq:phi}
    \phi_{\mu \nu} = P(f_\mu) \ \delta_{\mu \nu} \ / \ {\rm{t_{span}}},
\end{align}
with r.h.s. terms respectively corresponding to the power spectral density (PSD), the Kronecker delta and the time span of the data set. This approach corresponds to the Fourier-sum approach described in \citep{vanHaasteren2014}.

To summarize, we use a time-domain likelihood function with a parameterized covariance matrix that contains a PSD that is most often defined as a simple power-law described with two parameters: the amplitude $A$, usually set at the reference frequency of $1 \ {\rm yr}^{-1}$ and the spectral slope $\gamma$.

The Gaussian process signal in the time-domain $f(t)$ can be evaluated from Eq. \ref{eq:fweight} after estimating the weights $a$ with Eq. \ref{Eq:weightGP}. Let us now describe a method to estimate realizations of $a$, where we perform a Maximum Likelihood Estimation on Eq. \ref{Eq:weightGP} \citep{2013PhRvD..87j4021L}. Let us rewrite the logarithm of this equation in a matrix notation, 

\begin{align} 
    {\rm log} \left[ p(\delta t | a, GP) \right] = &- \frac{1}{2} (\delta t - a T)^{{\rm T}} N^{-1} (\delta t - a T) - \frac{1}{2} a^{{\rm T}} \phi^{-1} a \nonumber \\
    &- \frac{1}{2} \ \left[ (n+m) \ {\rm log} (2 \pi) + {\rm log} \left( {\rm det} \left( N \right) {\rm det} \left( \phi \right) \right) \right],
\end{align}
where the third term is just a constant number "${\rm cst}$", thus 
\begin{align} 
    {\rm log} \left[ p(\delta t | a, GP) \right] &= - \frac{1}{2}  \left[ \delta t^{{\rm T}} N^{-1} \delta t +  (a T)^{{\rm T}} N^{-1} (aT) + a^{{\rm T}} \phi^{-1} a \right] \nonumber\\ 
    & \ + (T^{{\rm T}} N^{-1} \delta t)^{{\rm T}} a + {\rm cst} \nonumber \\
    &= - \frac{1}{2}  \left[ \delta t^{{\rm T}} N^{-1} \delta t +  a^{\rm T} (T^{{\rm T}} N^{-1} T + \phi^{-1}) a \right] \nonumber \\
    & \ + (T^{{\rm T}} N^{-1} \delta t)^{{\rm T}} a + {\rm cst}.
\end{align}

Let us now write its derivative over the weights,
\begin{align} 
    \frac{\partial \ {\rm log} \left[ p(\delta t | a, GP) \right]}{\partial a} = (T^{{\rm T}} N^{-1} T + \phi^{-1}) a - (T^{{\rm T}} N^{-1} \delta t)^{{\rm T}},
\end{align}
and finally obtain the maximum likelihood vector of weights $\hat{a}$,

\begin{align} 
    \hat{a} = (T^{{\rm T}} N^{-1} T + \phi^{-1})^{-1} \ (T^{{\rm T}} N^{-1} \delta t)^{{\rm T}},
\end{align}
where we evaluate $\phi$ using Eq. \ref{eq:phi}, for which $P(f)$ is computed by using posterior distributions of our model parameters (e.g., $A$ and $\gamma$ if defined as a power-law) evaluated from a Bayesian analysis. Furthermore, $N$ is computed after applying white noise parameters (e.g., EFAC, EQUAD) with values taken from the same posterior distributions. This way, we evaluate time-domain realizations that are marginalized over other processes included during the Bayesian analysis.\\

\section{Achromatic noise recoveries with DMX and DM GP}
\label{Appendix: B}

In this section we report on how \textbf{DMX} and \textbf{DM GP} methods are able to recover achromatic noise parameters from our simulations.

\subsection{White Noise}

\begin{figure}
    \centering
    \includegraphics[width=0.6\linewidth, height=0.4\textheight, keepaspectratio]{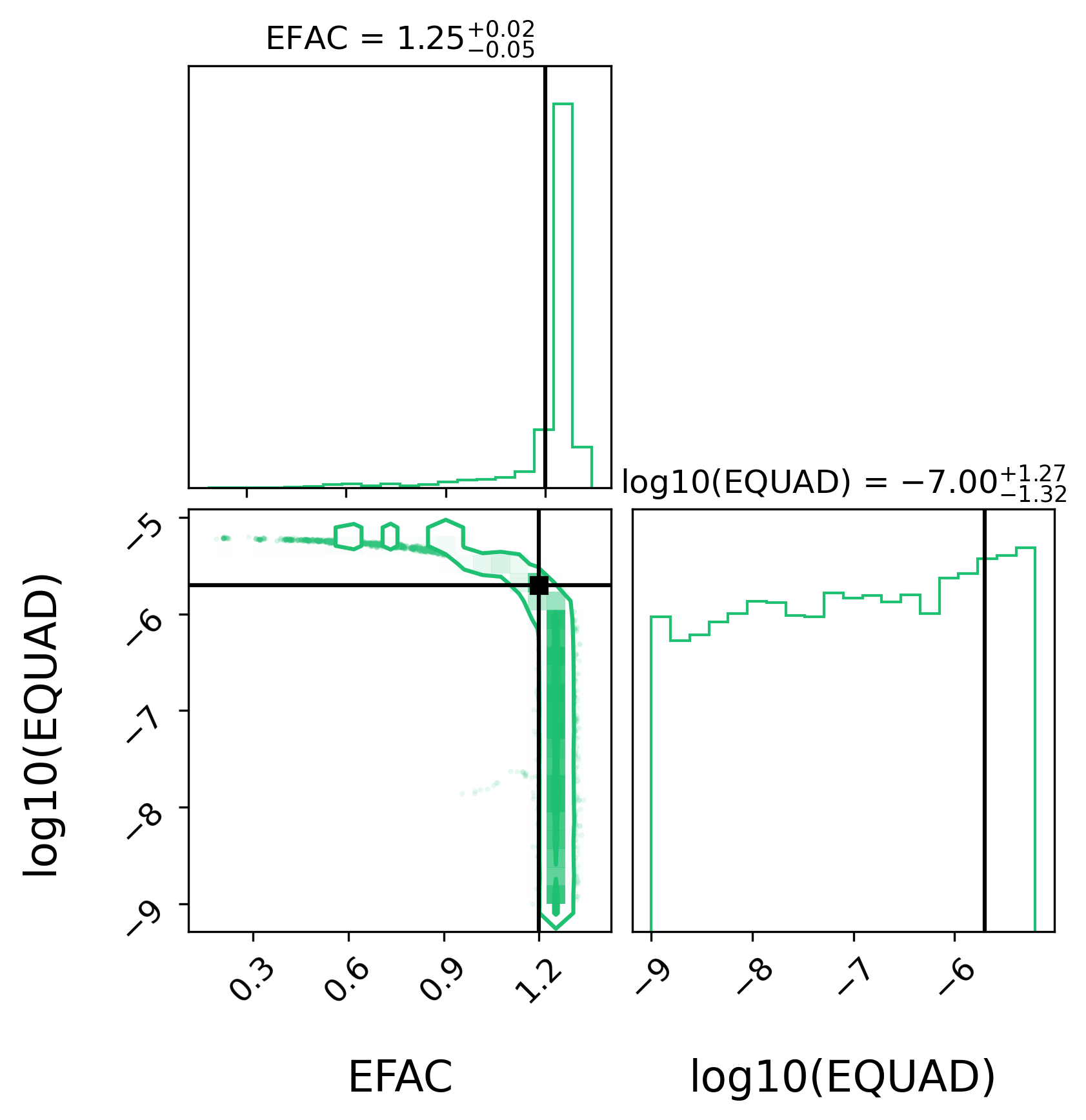}
    \caption{Corner plot of WN parameters using \textbf{DM GP} for an individual realization (see Section~\ref{sec:2}) with injected noise parameters: $A_{DM}=-13.3$, $\gamma_{DM}=-2.7$, $A_{RN}=-13.6$ and $\gamma_{RN}=-3.7$.
    Black lines correspond to the injected values.}
    \label{fig: WN posterior dm gp}
\end{figure}

\begin{figure}
    \centering
    \includegraphics[width=0.6\linewidth, height=0.4\textheight, keepaspectratio]{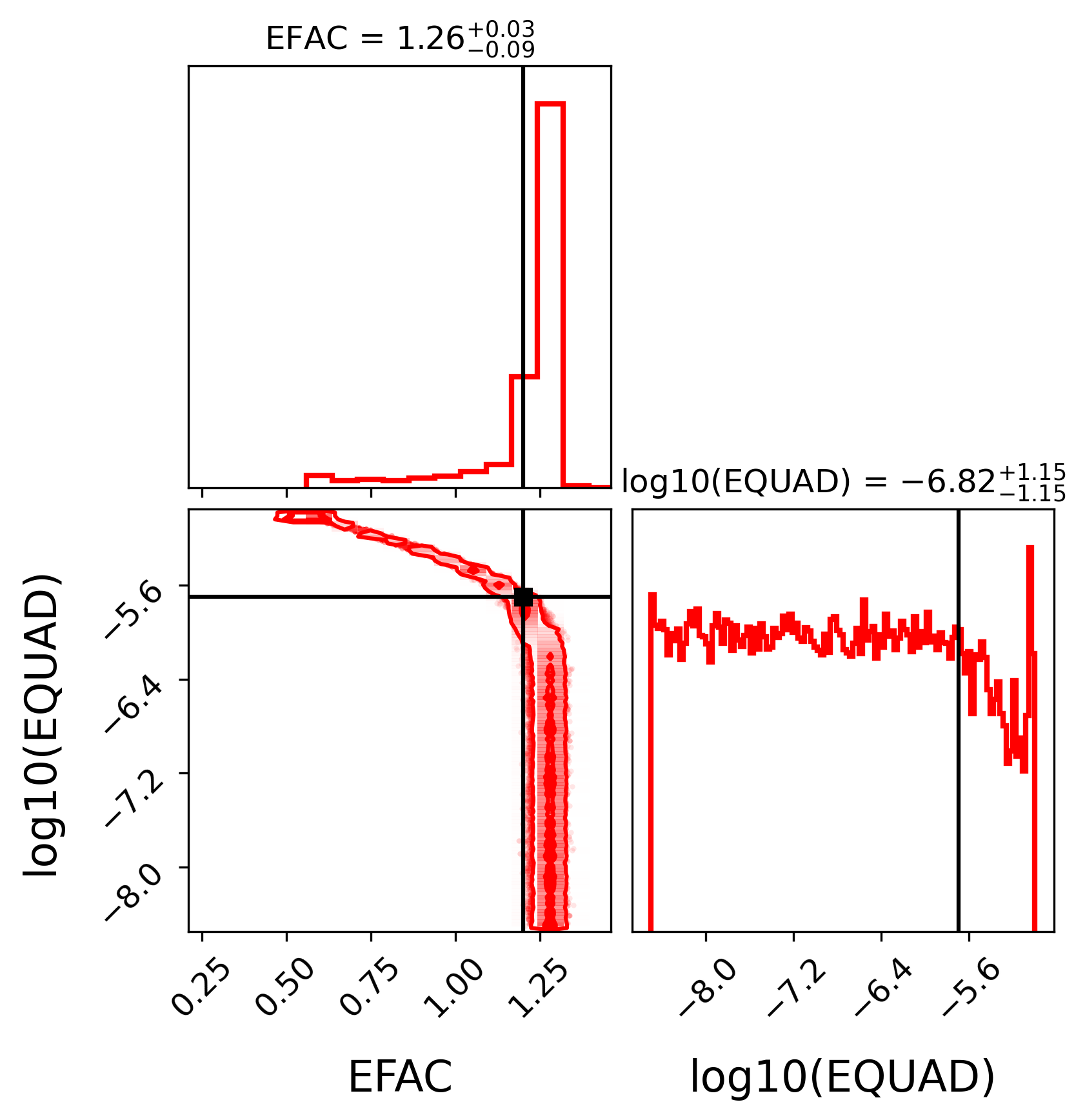}
    \caption{Corner plot of WN parameters using \textbf{DMX} for an individual realization (see Section~\ref{sec:2}) with injected noise parameters: $A_{DM}=-13.3$, $\gamma_{DM}=-2.7$, $A_{RN}=-13.6$ and $\gamma_{RN}=-3.7$.
    Black lines correspond to the injected values.}
    \label{fig: WN posterior dmx}
\end{figure}

WN parameters are inferred during the \textbf{DMX} and \textbf{DM GP} analysis through Equation~\eqref{eq: WN} in the \texttt{enterprise} software suite.
Since in our simulations the ToA uncertainties are all the same, the EFAC and EQUAD parameters are strongly correlated as shown in Figures~\ref{fig: WN posterior dm gp}~and~\ref{fig: WN posterior dmx}.
%
    Overall, the EFAC parameters are always well constrained and consistent with the injected values for both \textbf{DM GP} and \textbf{DMX}.
    On the other hand, the EQUADs show a flat posterior (meaning that they are totally unconstrained) but still consistent with the injection. 
    Moreover, \textbf{DM GP} tends to present a small peak at higher EQUADs, while \textbf{DMX} does not.
    In order to disentangle EFAC and EQUAD parameters, simulations with variable ToA uncertainties are required.

\subsection{Achromatic red noise}
\begin{figure}
        \centering
        \includegraphics[width=\linewidth]{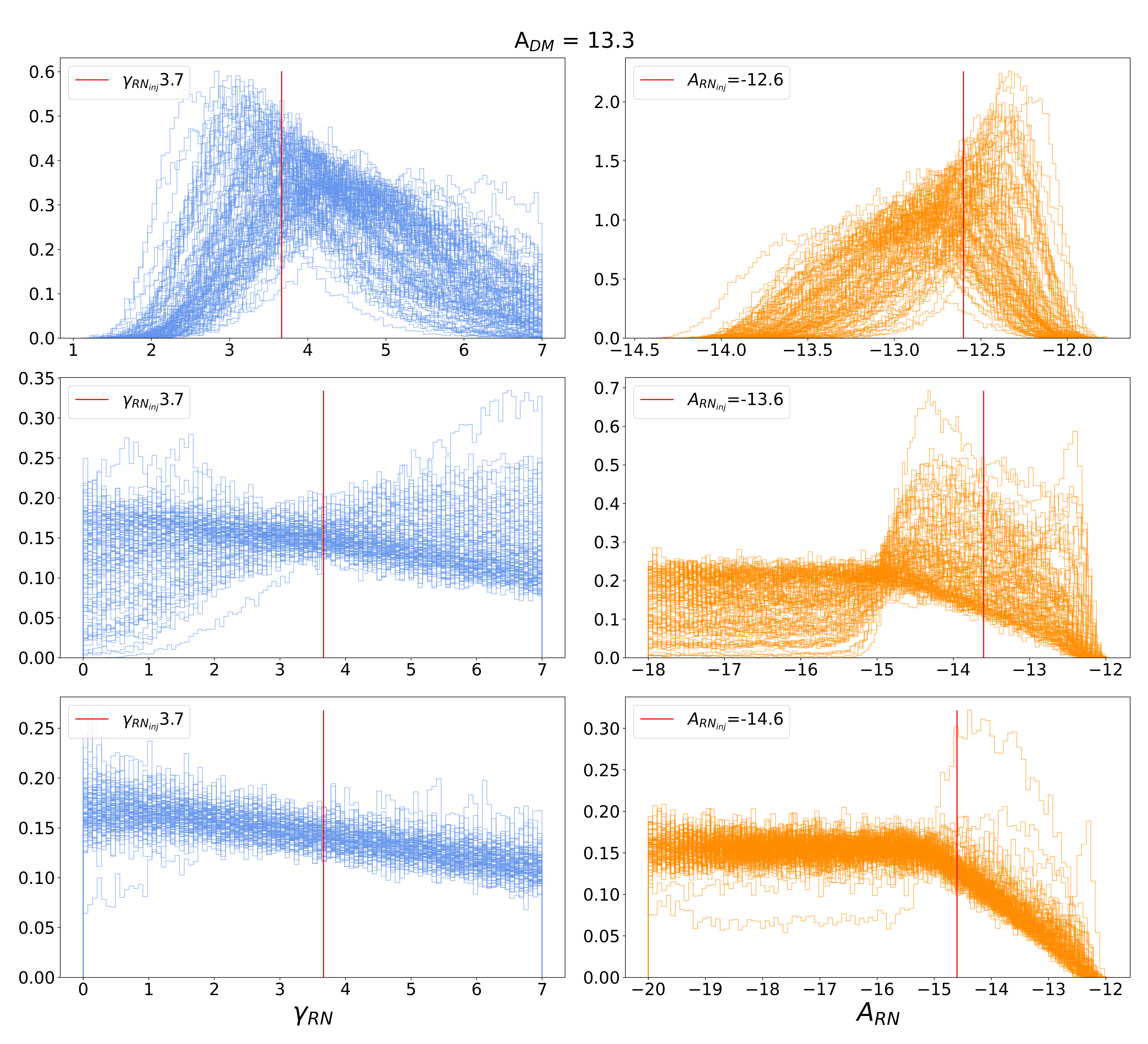}
        \caption{Posterior distributions of RN parameters using \textbf{DM GP} for 100 realizations (see Section~\ref{sec:2}) with injected DM noise parameters: $A_{DM}=-13.3$, $\gamma_{DM}=-2.7$.
        On the left-hand column there is the spectral index $\gamma_{RN}$; on the right-hand column there is the RN amplitude $A_{RN}$.
        Each row correspond to a different injected $A_{RN}$.
        Red vertical lines report the injected values.}
        \label{fig: RN pars GP}
    \end{figure}

\begin{figure}
        \centering
        \includegraphics[width=\linewidth]{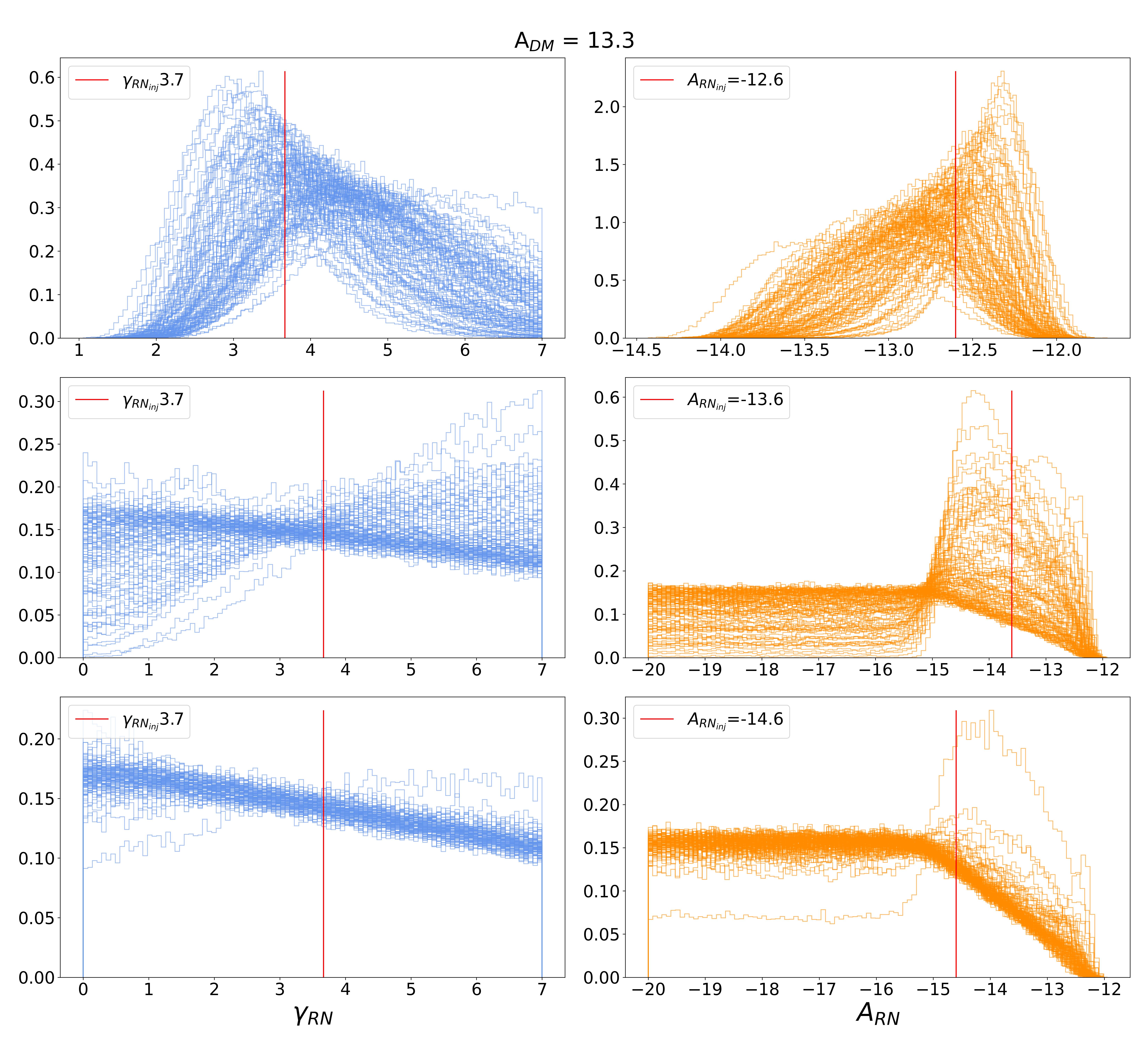}
        \caption{Posterior distributions of RN parameters using \textbf{DMX} for 100 realizations (see Section~\ref{sec:2}) with injected DM noise parameters: $A_{DM}=-13.3$, $\gamma_{DM}=-2.7$.
        On th left-hand column there is the spectral index $\gamma_{RN}$; on the right-hand column there is the RN amplitude $A_{RN}$.
        Each row correspond to a different injected $A_{RN}$.
        Red vertical lines report the injected values.}
        \label{fig: RN pars DMX}
    \end{figure}

    In Figures~\ref{fig: RN pars GP}~and~\ref{fig: RN pars DMX} we report the 100 posterior distributions of RN parameters obtained from the simulations described in section \ref{sec:2}.
    We show the only fiducial case with $A_{DM}=-13.3$ and varying $A_{RN}$.
    \texttt{Enterprise} can well constrain the RN parameters alongside \textbf{DM GP} and \textbf{DMX} as long as the power spectrum of the injected RN is above the WN floor (i.e., when  $A_{RN}=-12.6$ and barely for $A_{RN}=-13.6$. See also Figure~\ref{fig: sim injection plot}).
    When this does not happen, the posterior distributions of $A_{RN}$ and $\gamma_{RN}$ are flat and unconstrained.
    In all of the cases the recoveries are consistent with the injected values.
    This result is also confirmed when we reduce the ToA uncertainties to $1$ or $0.1~\mathrm{\mu}s$, hence decreasing the WN power level.
    In these cases the RN parameters are well constrained and consistent with the injection even for lower $A_{RN}$.

\end{appendix}

%
%

\end{document}